\documentclass[twocolumn,usenatbib,useAMS]{mnras}

\PassOptionsToPackage{dvipsnames,svgnames,x11names}{xcolor}
\usepackage{xparse,soul}

\makeatletter
\NewDocumentCommand{\sotwo}{O{red}O{black}+m}
    {%
        \begingroup
        \color{#1}%
        \setul{-.5ex}{.4pt}%
        \def\SOUL@uleverysyllable{%
            \rlap{%
                \color{#2}\the\SOUL@syllable
                \SOUL@setkern\SOUL@charkern}%
            \SOUL@ulunderline{%
                \phantom{\the\SOUL@syllable}}%
        }%
        \ul{#3}%
        \endgroup
    }
\makeatother

\usepackage{graphicx}	
\usepackage{amsmath}	
\usepackage{amssymb}	
\usepackage{multicol} 
\usepackage{multirow}
\usepackage{bm}	
\usepackage{pdflscape}
\usepackage{caption}
\usepackage{subfigure}
\usepackage{siunitx}
\usepackage{color}

\usepackage[T1]{fontenc}
\usepackage{ae,aecompl}

\usepackage{newtxtext,newtxmath}

\title[How dark are filaments in the cosmic web?]{How dark are filaments in the cosmic web?}

\author[T. Yang et. al.]{
Tianyi Yang$^{1,2,3}$\thanks{E-mail: t65yang@uwaterloo.ca},
Michael J. Hudson$^{1,2,3}$\thanks{E-mail: mike.hudson@uwaterloo.ca},
Niayesh Afshordi$^{1,2,3}$\thanks{E-mail: nafshordi@pitp.ca}
\\
$^{1}$Department of Physics and Astronomy, University of Waterloo, Waterloo, ON, N2L 3G1, Canada\\
$^{2}$Waterloo Centre for Astrophysics, University of Waterloo, Waterloo, ON, N2L 3G1, Canada\\
$^{3}$Perimeter Institute of Theoretical Physics, 31 Caroline St. N., Waterloo, ON, N2L 2Y5, Canada
}
\date{Accepted XXX. Received YYY; in original form ZZZ}

\pubyear{2020}

\begin{document}
\label{firstpage}
\pagerange{\pageref{firstpage}--\pageref{lastpage}}
\maketitle

\begin{abstract}
The cold dark matter model predicts that dark matter haloes are connected by filaments. Direct measurements of the masses and structure of these filaments are difficult, but recently several studies have detected these dark-matter-dominated filaments using weak lensing. Here we study the efficiency of galaxy formation within the filaments by measuring their total mass-to-light ratios and stellar mass fractions. Specifically, we stack pairs of Luminous Red Galaxies (LRGs) with a typical separation on the sky of $8 h^{-1}$ Mpc. We stack background galaxy shapes around pairs to obtain mass maps through weak lensing, and we stack galaxies from the Sloan Digital Sky Survey (SDSS) to obtain maps of light and stellar mass. To isolate the signal from the filament, we construct two matched catalogues of physical and non-physical (projected) LRG pairs, with the same distributions of redshift and separation. We then subtract the two stacked maps. Using LRG pair samples from the BOSS survey at two different redshifts, we find that the evolution of the mass in filament is consistent with the predictions from perturbation theory. The filaments are not entirely dark: their mass-to-light ratios ($M/L = 351\pm137$ in solar units in the $r$-band) and stellar mass fractions ($M_{\rm stellar}/M = 0.0073\pm0.0030$) are consistent with the cosmic values (and with their redshift evolutions). 
\end{abstract}

\begin{keywords}
 cosmology: dark matter -- cosmology: large-scale structure of Universe -- galaxies: elliptical and lenticular, cD -- gravitational lensing: weak
\end{keywords}

\section{Introduction}\label{sec::intro}

One of the predictions of the $\Lambda$ cold dark matter ($\Lambda$CDM) model is the formation of web-like structures on large scales, the so-called ``Cosmic Web'' \citep{Bond_web}. These can be seen in N-body simulations \citep[e.g.][]{Mil_simulation, Illustris_sim}, or observationally, though the maps of galaxy distribution probed by redshift surveys such as 2dFGRS \citep{2dFGRS_fil}, SDSS \citep{SDSS_fil}, 2MASS \citep{2MASS_fil}, or by VIPERS \citep{VIPER_fil} at higher redshift. One of the most prominent features of the Cosmic Web are the filaments: moderately overdense regions that connect two massive dark matter haloes. 
While filamentary structures have been traced by galaxies, their dark matter properties remain poorly understood observationally. N-body simulations can be used, however, to predict the properties of filaments. Filament candidates can be identified by extracting pairs of massive haloes in high-resolution N-body simulations \citep[e.g.][]{Intercluster_filaments_in_Universe, Statistical_properties_of_filaments_in_weak_gravitational_lensing}. Other methods have also been developed for detecting and mapping the filamentary structures, and  all methods confirm that the most massive haloes reside at the intersection of filaments (the so-called knots) and that, apart from the knots, filaments are the most massive and densest structures \citep[see][for a brief review.]{Tracing_the_cosmic_web}.

Weak gravitational lensing (WL) is the most promising way to measure the dark matter content of filaments. In WL, a foreground matter distorts (``shears'') the shapes of background galaxies. The \citet{Kaiser_Squires_lensing} algorithm can be used to convert the measured shear field into a map of the projected mass density. Using this principle, \citet{Abell_cluster_lensing_detection} reported a filament connecting Abell 222 and Abell 223 supercluster. \citet{A_weak_lensing_mass_reconstruction_filament_MACS} found a 3$\sigma$ filament detection in the massive galaxy cluster field MACS J0717.5+3745. Less massive filaments, such as those between group-mass haloes, have too low a density contrast to detect individually, and so a measurement is only possible for a stacked ensemble of filaments. One way to identify filaments is to use pairs of luminous red galaxies (LRG). LRGs are red early-type galaxies that are selected based on their colours and magnitudes \citep{LRG_selected, Dawson_paper}. Most LRGs reside in the centres of group-mass ($\sim 10^{13} M_{\odot}$) dark matter haloes \citep{LRG_good_aporxy_rich_group, LRG_halo_center}. \citet{clampitt_fil_detection} reported a $4.5\sigma$\ detection of filaments after stacking 135,000 LRG pairs selected from SDSS DR7 \citep{SDSS_DR7_LRG}. \citet{Seth_filament_paper} mapped the filament by stacking $\sim$ 20,000 LRG pairs selected from Baryon Oscillation Spectroscopic Survey \citep[][hereafter BOSS]{Eisenstein_boss_paper, Dawson_paper} and the background source galaxy ellipticities from CFHTLens \citep{ellipticity_CFHTLens, CFHTLens_general_Erben}. Subsequently, WL detections of the stacked filaments between LRG pairs have been reported by a number of authors \citep[e.g.][]{Kondo_stacking_fil, new_filament_paper, cmb_filament}.

While there have now been both studies of the luminous content of filaments and of their dark matter content, to our knowledge there have been no attempts to measure both simultaneously in a consistent way and interpret the results therefrom. One way to quantify this comparison is the mass-to-light ratio (or the stellar-to-total mass ratio) of filaments. This is of interest for several reasons. It is well known, both from abundance matching \citep{m_L_relation_Marinoni_hudson, mass_light_ratio_halo, Behroozi_abundance_matching_paper_1, Behroozi_abundance_matching_paper_2, relation_galaxy_dmhalo_review} and from weak lensing, dynamics and X-ray studies \citep{Parker_M_L,  More_satellite_dynamics_SDSS, Sheldon_M_L, lensing_source_galaxy_selection, Kravtsov_2018_SF_efficiency}, that the mass-to-light and the stellar-to-total-mass ratios depends on halo mass. Specifically, the mass-to-light ratio is a minimum (and the stellar-to-total-mass ratio is a maximum) for haloes with masses around $\gtrsim 10^{12} M_{\odot}$. Since the abundance of haloes, i.e.\ the halo mass function, also depends on environment, being different in filamentary environments than in knots \citep{Tracing_the_cosmic_web}, the mass-to-light ratio in a filament is likely to reflect this dependence and therefore may be different from the global mean. Of course, it is also possible that the filament environment affects the mass-to-light ratio of haloes at fixed halo mass. The goal of this paper is to measure the mass-to-light and stellar-to-total-mass ratios, as a first step towards understanding these effects.

This paper is structured as follows: in Section \ref{sec::data}, we discuss the data selection of LRG pairs and source galaxies, where LRGs are selected from BOSS and ellipticities of source background galaxies are taken from CFHTLenS. In Section \ref{sec::mass_map}, using weak lensing data, we reproduce the 2D projected mass maps for stacked filaments following the methodology presented in \citet{Seth_filament_paper}. In Section \ref{sec:: light_map}, adopting the same LRG pair selection, we then stack the galaxy light and calculate the average luminosity of filaments. We present the resulting 2D projected luminosity density map and measured luminosity of stacked filaments in Section \ref{sssec::observed_light_map}. We compare our $M/L$ measurements for filaments with other literature in Section \ref{sec::discussion}. In Section \ref{sec::conclusion}, we summarise our results and draw conclusions. Throughout this work, we assume a $\Lambda$CDM flat cosmological model and the relevant cosmological parameters are adopted as follow: $\Omega_{m,0}$ = 0.3, $\Omega_{\Lambda,0}$= 0.7 and $h \equiv$ $H_0$/(100 km $\textrm{s}^{\textrm{-1}}$ $\textrm{Mpc}^{\textrm{-1}}$) = 0.7.

\section{Data}\label{sec::data}

Filaments are the bridges between galaxy groups and clusters, but only pairs that are relatively close to each other are expected be connected by filaments. As LRGs are considered to be good proxies for rich groups \citep{LRG_good_aporxy_rich_group}, we first focus on the selection of LRGs in Section \ref{ssec::data_selection_LRG}. Pairs of LRGs that are close in redshift are referred to as \emph{physical} LRG pairs, and these are used to identify filaments. Furthermore, to isolate the filament signal, we also need a control sample of pairs that are not physically connected (close in redshift), which are referred to as \emph{non-physical} (projected) pairs. The criteria for the pair selection are outlined in Section \ref{ssec::data_selection_LRG_pairs}.

\subsection{LRG selection}\label{ssec::data_selection_LRG}

To study the filaments between physical LRG pairs, accurate LRG redshifts are necessary. Here, we use the spectroscopic redshifts provided by SDSS-III/BOSS to select LRGs. BOSS is a spectroscopic survey covering over 10,000 square degrees derived from the SDSS imaging, including the spectra and redshift measurements for 1.5 million galaxies with redshift $z < 0.7$. A full description and summary of the survey are provided in \citet{Eisenstein_boss_paper} and \citet{Dawson_paper}.

The weak lensing signal from the filament connecting a single LRG pair is weak and very noisy, so we stack LRG pairs together to enhance the significance of detection. Therefore, we extracted all object redshifts from 0.15 to 0.7 that are flagged with SourceType = ``LRG'' and Z$\_$NOQSO>0. In order to investigate how the properties of filaments evolve with redshift, the whole LRG sample is further divided into two independent redshift bins: LOWZ and CMASS. In BOSS, LOWZ (low redshift) aims to selecting galaxies at lower redshift with coverage $0.15<z<0.43$ and CMASS (constant mass) is designed to select galaxies within $0.43<z<0.7$. In SDSS DR14, LRGs which satisfy with the LOWZ or CMASS criteria are flagged with ``GAL$\_$LOZ'' or ``GAL$\_$CMASS$\_$ALL'' respectively. From now on, we shall have three samples available for further analysis: the whole BOSS sample including all LRGs from $z$ = 0.15 to 0.7, and the LOWZ and CMASS subsamples. 

\subsection{LRG physical and non-physical pairs}
\label{ssec::data_selection_LRG_pairs}

For consistency with \citet{clampitt_fil_detection} and \citet{Seth_filament_paper}, physical LRG pairs are constructed by selecting pairs that have redshift separation $\Delta{z_{\textrm{sep}}}<0.002$ and projected 2D $R$ separation within $6h^{-1} \textrm{Mpc} \leqslant R_{\textrm{sep}} \leqslant 10 h^{-1} \textrm{Mpc}$. They are physically ``real'' pairs as they are not only close in projection, but are also close in redshift space. In other words, two LRG haloes with such a small redshift separation are expected to be connected by filament.

To isolate the mass associated with the filament (as opposed to mass that may be clustered with LRGs), we also define a catalogue of non-physical LRG pairs. Non-physical LRG pairs, which are close in projection only, are selected by the same $R_{\textrm{sep}}$ projection criterion as for physical LRG pairs but with larger redshift separation: $0.033 \leqslant z_{\textrm{sep}} \leqslant 0.05$. Galaxy groups/clusters which have this amount of separation in the redshift space are unlikely to have filaments between them, so the signal from non-physical pairs only yields the contribution from two isolated LRG haloes. Therefore, the residual signal from the filament should remain after the subtraction between the stacked maps of physical and non-physical pairs.

\section{Lensing mass map}\label{sec::mass_map}

Following the procedures outlined in \citet{Seth_filament_paper}, in this section, we reproduce the 2D stacked surface mass density map of physical and non-physical LRG pairs using weak gravitational lensing. A major difference between our study and \citet{Seth_filament_paper} is that we also repeat the same analysis with BOSS LOWZ and CMASS subsamples separately. We begin with the LRG pair and lensing source galaxies selection in Section \ref{ssec::lensing_pair_selection} and \ref{ssec::data_selection_source_galaxies_mass}. We present the procedures for the generation of lensing map in Section \ref{ssec::lensing_map_procedures}. Then we summarise and discuss the lensing results in Section \ref{ssec::lensing_results}. 

\subsection{Lensing pair selection}\label{ssec::lensing_pair_selection}

The physical and non-physical pairs are selected based on the criteria outlined in \ref{ssec::data_selection_LRG_pairs}. However, some extra position cuts need to be included. The lensing map is derived from ellipticity measurements from CFHTLens survey, so only LRGs in the regions of sky covered by CFHTLens survey are selected. This yields 15,254 LRG physical pairs in total, with a mean redshift <$z_{\textrm{pair}}$> $\sim$ 0.47 and a mean 2D projected distance <$R_{\textrm{sep}}$> $\sim$ 8.10$h^{-1}$ Mpc. For the LOWZ sample, there are 2,752 LRG physical pairs with <$z_{\textrm{pair}}$> $\sim$ 0.30 and <$R_{\textrm{sep}}$> $\sim$ 8.10 $h^{-1}$ Mpc, and for the CMASS sample, there are 12,497 LRG physical pairs with <$z_{\textrm{pair}}$> $\sim$ 0.52 and <$R_{\textrm{sep}}$> $\sim$ 8.04 $h^{-1}$ Mpc.

\subsection{Lensing source galaxies}
\label{ssec::data_selection_source_galaxies_mass}
The weak lensing source galaxies are selected from the CFHTLenS catalogues. CFHTLenS is derived from the Wide component of the Canada-France-Hawaii Telescope Legacy Survey (CFHTLS) and covers 154 square degrees including the photometric information from five optical bands ($u_{\ast}$, $g'$, $r'$, $i'$, $z'$) in four patches: $W_{1} - W_{4}$, where $W1, W3$ and $W4$ have substantial overlap with BOSS over an area of approximately 100 $\rm deg^2$ \citep{CFHT_lens_region_cut}. The galaxy shape measurements are from the $\textit{lensfit}$ algorithm and the photometric redshifts are obtained using the Bayesian Photometric Redshift code with a scatter of $\sigma_{z} \sim 0.04(1+z)$. General information about the data reduction and the survey design can be found in \citet{CFHTLens_general_Heymans} and \citet{CFHTLens_general_Erben}, and a detailed description of the determination of the ellipticity and redshift measurements are presented in \citet{ellipticity_CFHTLens} and \citet{CFHTLens_redshift} respectively.

Based on the criteria outlined in \citet{lensing_source_galaxy_selection}, we select all source galaxies from unmasked regions with reliable photometric redshift measurements (0.2 < $z_{\textrm{p}}$ < 1.3) and with FITCLASS=0. Only galaxies with weight $w > 0$ are used. This selection yields approximately $5.6 \times 10^{6}$ background sources. We use source galaxies that have photometric redshifts greater than 0.1 of that of the LRG pair, in order to avoid contamination by intrinsically-aligned ``source'' galaxies within the filaments themselves.

\subsection{Method}\label{ssec::lensing_map_procedures}

As is the case with galaxy-galaxy lensing, it is necessary to stack LRG pairs to extract a statistically significant signal. Lensing by LRG pairs, however, is more complicated than lensing by single galaxies because the pairs are not circularly symmetric. Furthermore, LRG pairs have random orientation angles on the sky and each galaxy pair has its own projected separation. Therefore we need a coordinate transformation to normalise the position of each LRG pair in order to compare and stack the lensing signals. To achieve this, we follow the procedure outlined in \citet{Seth_filament_paper}. First, all LRG pairs are rotated with respect to their centres such that the two LRGs lie along the $x$-axis in the new coordinate system. Then, pairs are rescaled by their 2D projected separation and the coordinate of individual LRG is transformed to ($x_{\textrm{L}}$, $y_{\textrm{L}}$) = (-0.5, 0) and ($x_{\textrm{R}}$, $y_{\textrm{R}}$) = (0.5, 0), where L and R denote left and right. Finally, all LRG pairs are stacked together. The coordinate transformation of the source galaxies follows the same procedure. Following this procedure, the shear map can be generated by stacking the ellipticities of the background source galaxies.

The resulting shear map, however, is difficult to interpret. Therefore, we convert the shear map into a convergence map, which is directly proportional to the surface mass density of the field, using the method of \citet{Kaiser_Squires_lensing}. With the convergence map, we can convert it to surface mass density based on the definition of convergence:
\begin{equation}\label{eqn::convergence_definition}
    \Sigma = \kappa \overline{\Sigma}_{\textrm{crit}},
\end{equation}
where $\overline{\Sigma}_{\textrm{crit}}$ is the ensemble average. This is calculated using the following formula:
\begin{equation}\label{eqn:average_critical_sigma}
    \overline{\Sigma}_{\textrm{crit}} = \frac{\sum_{l}\sum_{s}\Sigma_{\textrm{crit},ls} W_{ls}} {\sum_{l}\sum_{s} W_{ls}},
\end{equation}
where $\Sigma_{\textrm{crit},ls}$ is the critical density for a given lens-source pair, $W_{ls} = w_{s}\Sigma_{\textrm{crit},ls}^{-2}$, $w_{s}$ is the lensfit weight of a source. For the whole LRG sample, the mean critical density is $\overline{\Sigma}_{\rm crit} = 3471 M_{\odot}/\rm pc^2$. For the LOWZ sample, we have $\overline{\Sigma}_{\rm crit} = 3320 M_{\odot}/\rm pc^2$, and for the CMASS sample, $\overline{\Sigma}_{\rm crit} ~\textrm{is found to be} ~3565 M_{\odot}/\rm pc^2$.

\begin{figure*}
    \begin{minipage}[b]{0.5\textwidth}
        \centering
        \includegraphics[width=\linewidth]{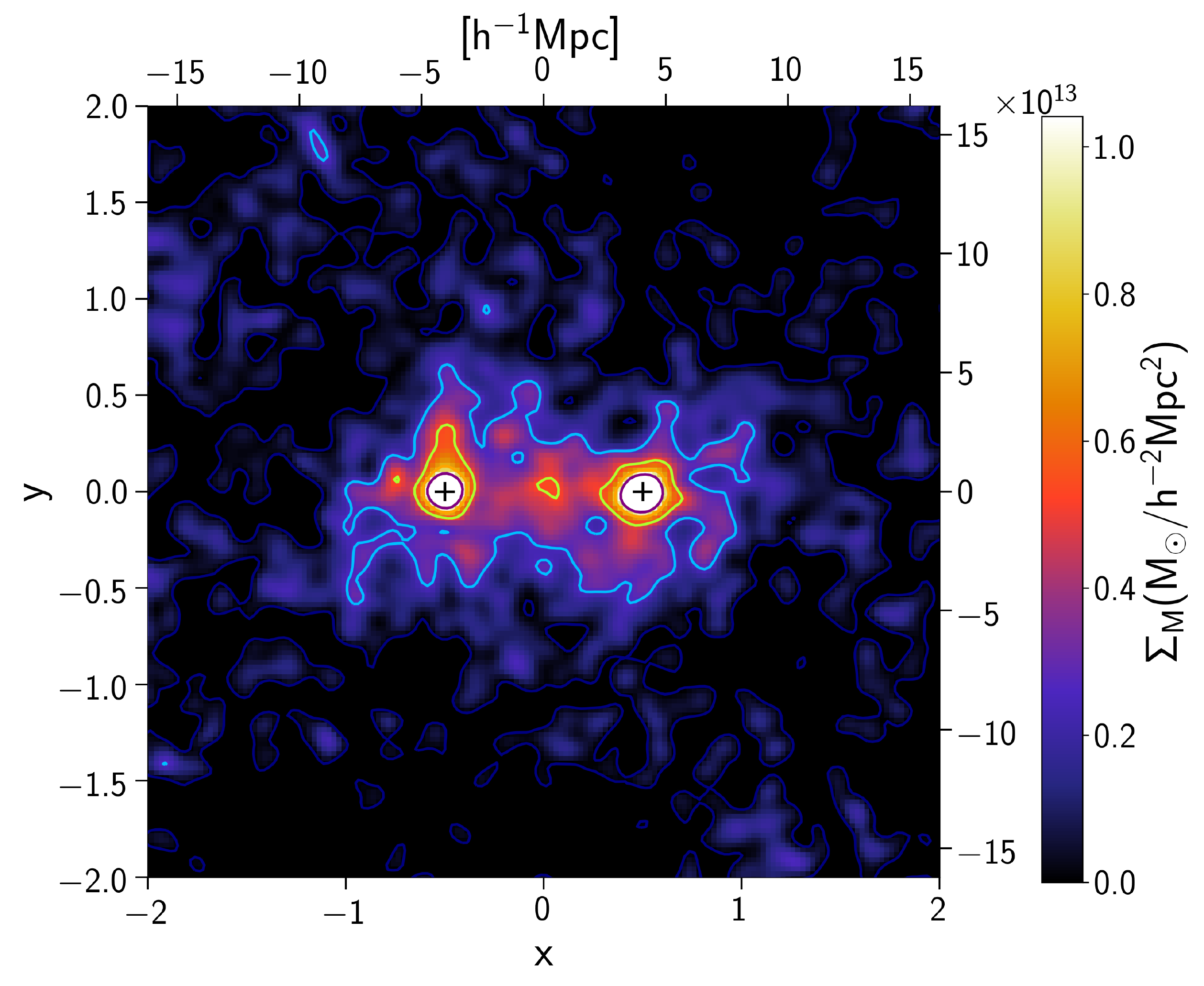}
    \end{minipage}\begin{minipage}[b]{0.5\textwidth}
        \centering
        \includegraphics[width=\linewidth]{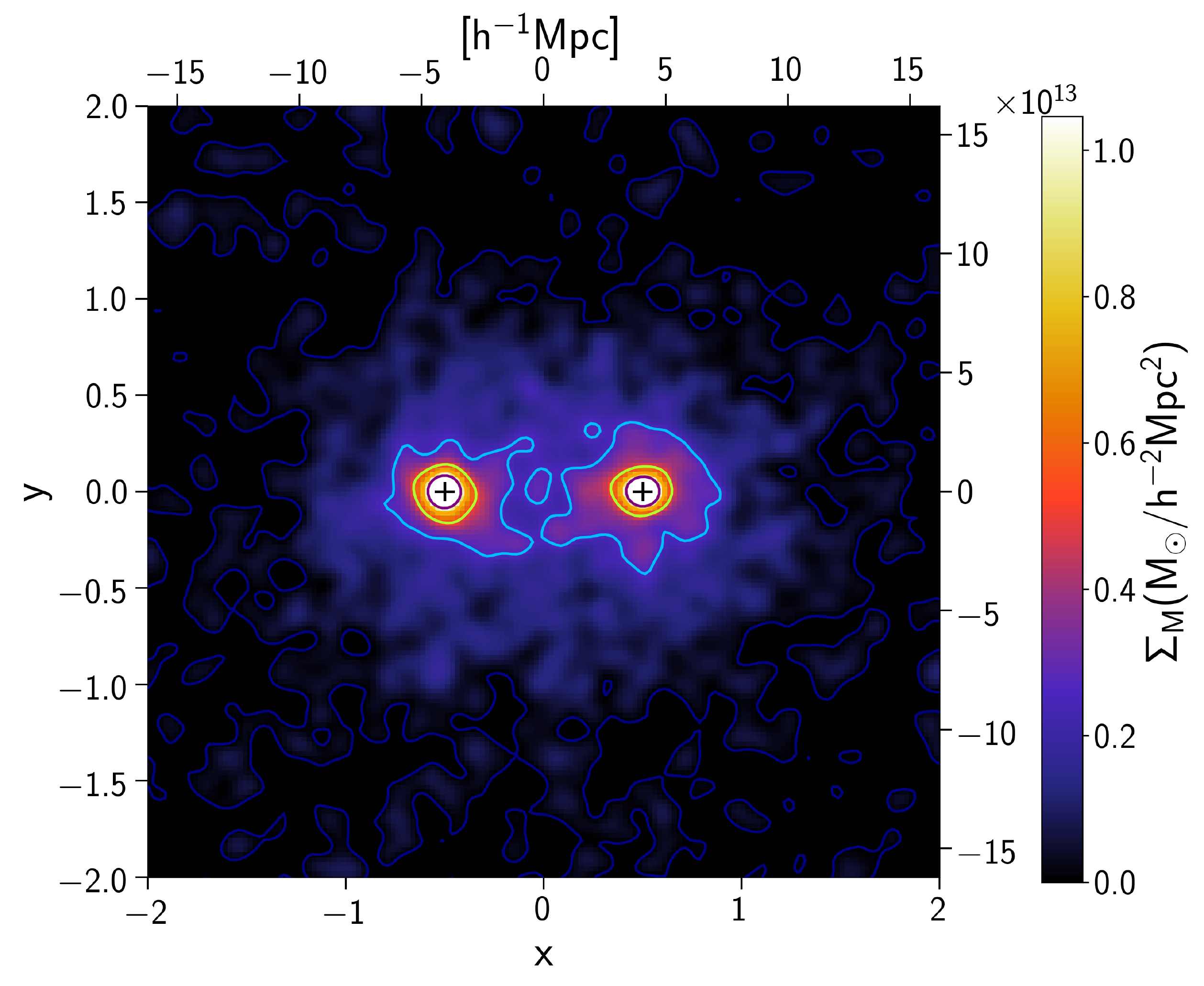}
    \end{minipage}\\
    \begin{minipage}[b]{0.5\textwidth}
        \centering
        \includegraphics[width=\linewidth]{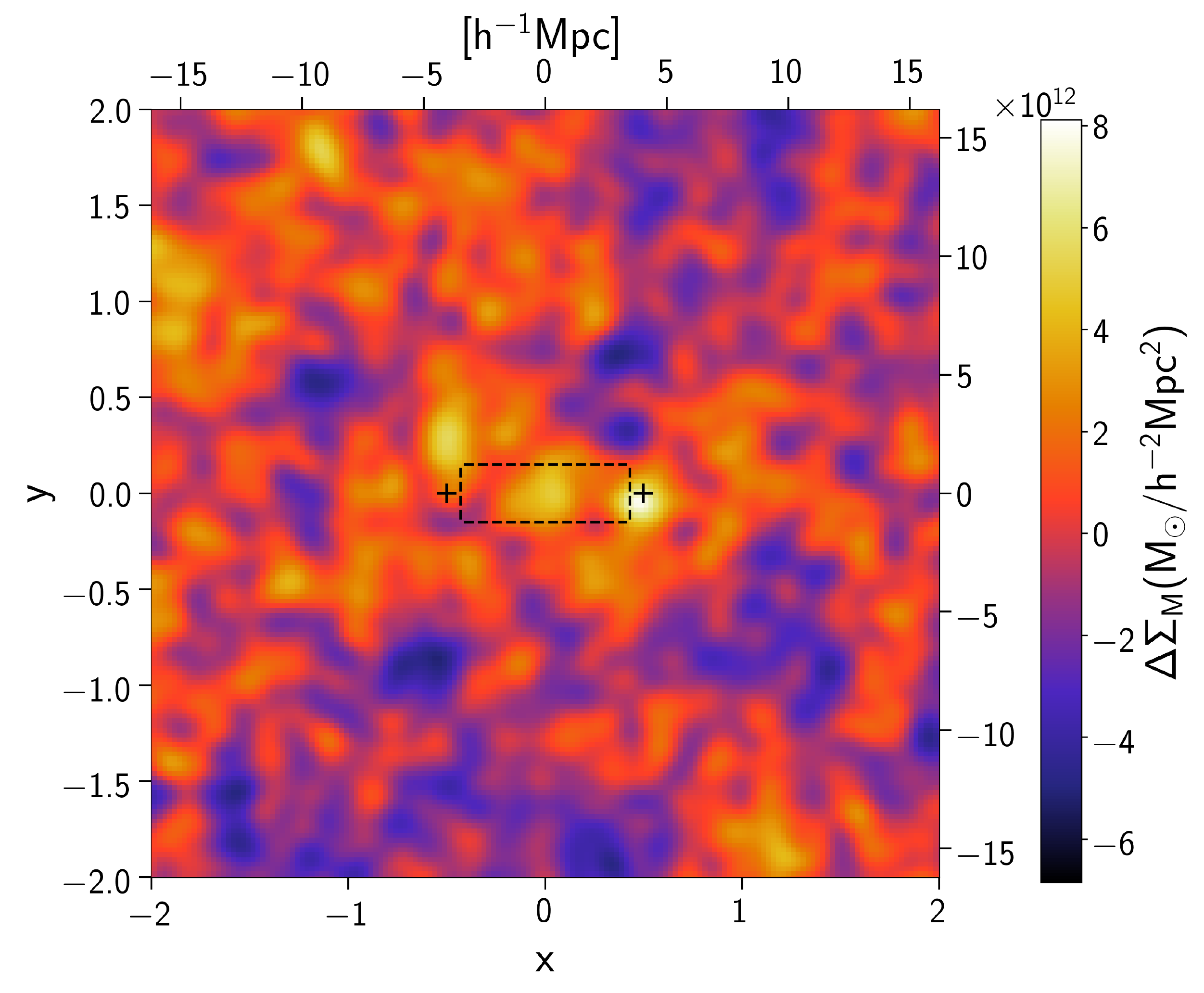}
    \end{minipage}\begin{minipage}[b]{0.5\textwidth}
        \centering
        \includegraphics[width=0.9\linewidth]{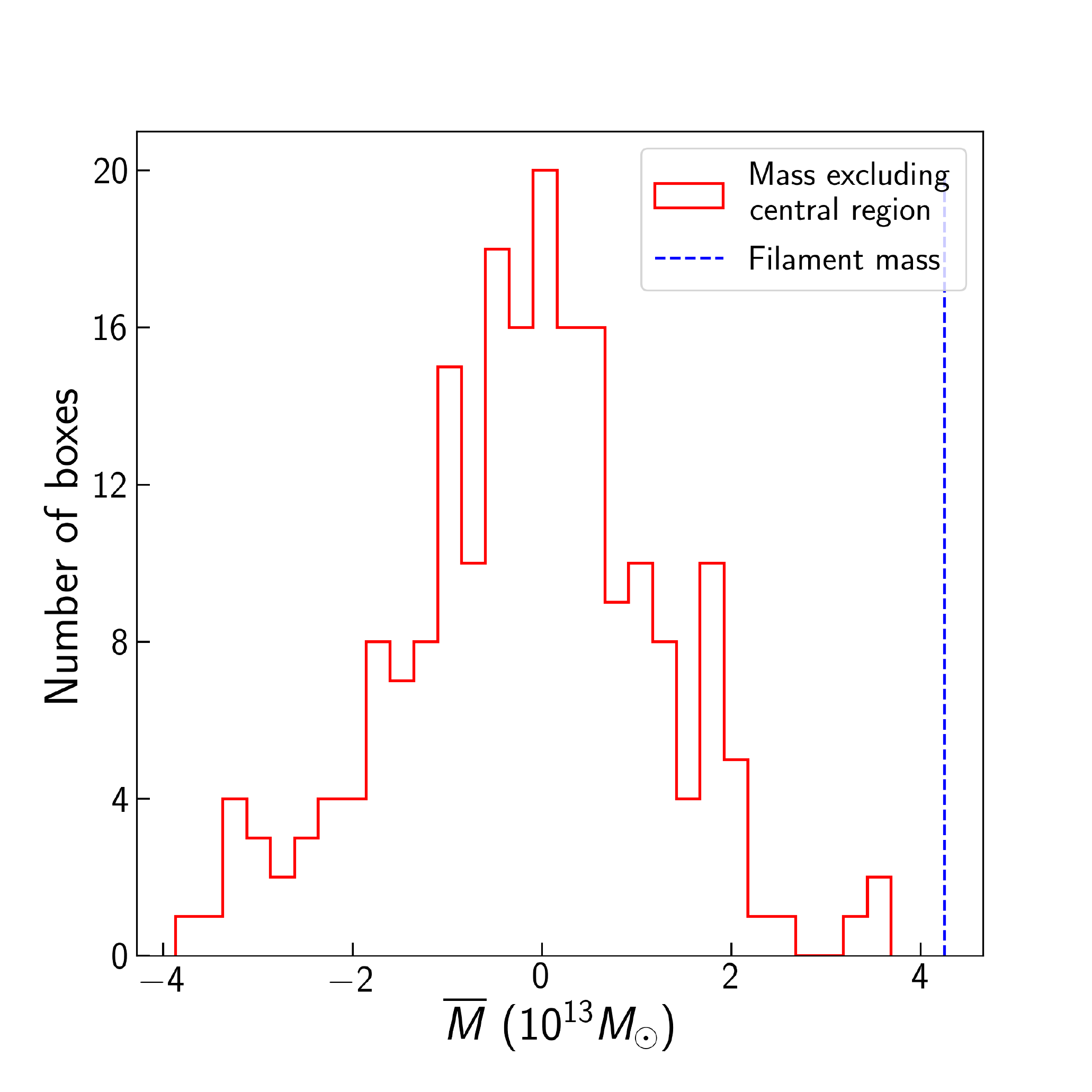}
    \end{minipage}%
\caption{\textit{Top panel}: the surface mass density map ($\Sigma_M$) obtained from the shear map using the \citet{Kaiser_Squires_lensing} algorithm. The locations of LRG haloes in the standardized coordinate system are marked as two ``+'' signs. A Gaussian filter of smoothing scale 0.40 $h^{-1}\rm Mpc$ (0.05 in units of $x,y$) has been applied to the image for illustration. Note that the left and bottom coordinates are in rescaled units, while the top and right are in $h^{-1}$ Mpc. The colorbar shows the surface mass density in units of solar mass per $h^{-2}\textrm{Mpc}^2$. Top left:  map constructed using physical LRG pairs, where the central bridge-like filamentary structure is visible. Top right: $\Sigma_M$ map from  non-physical LRG pairs. \textit{Bottom left panel}: the resulting excess surface density map after the subtraction the $\Sigma_M$ map of the non-physical pairs from that of the physical pairs. The region which is used for the mass measurement of the filament is shown by the black dashed rectangular box. The size of the filament box is 0.86 $\times$ 0.30 $R_{\rm sep}$ \citep{Seth_filament_paper} in the new coordinate system, which corresponds to $\sim$ 7.0 $h^{-1} \textrm{Mpc}$ $\times$ 2.4 $h^{-1} \textrm{Mpc}$ in proper size. \textit{Bottom right panel}: histogram of the averaged masses within the same sized rectangular box as displayed in the bottom left panel. Specifically, boxes are placed on the excess convergence map over a region between $\pm$ 4 along the x axis and $\pm$ 4 along the y axis. A square centred at (0,0) with an area of 4 is excluded in order to avoid the central filament region.}
\label{lensing_plots}
\end{figure*}

\subsection{Lensing results}\label{ssec::lensing_results}
 
The projected surface mass density maps using all LRG samples are shown in the top panel of Figure \ref{lensing_plots}, with the left panel demonstrating the map reconstructed by using physical pairs connected by a central filament. The top right panel shows the projected surface mass density around non-physical pairs. In contrast with the physical pairs, there is no obvious filamentary structures between them. The excess projected surface mass density after subtraction is in the bottom panel with a rectangular box showing the region used for the mass measurement of the filaments. The proper size of the filament box is $\sim$ 7.0 $h^{-1} \textrm{Mpc}$ $\times$ 2.4 $h^{-1} \textrm{Mpc}$, following \citet{Seth_filament_paper}. The size of the box is the projected $R_{\rm sep}$ after averaging all LRG pairs, and it has excluded a region extending to 3 $r_{200}$ around the LRG haloes.
 
The uncertainty in the measured filament mass is mainly due to ``shape noise''. We quantify this by first defining the variance of the shear in a given pixel at location $(x,y)$ in the new coordinate system, $\gamma(x,y)$, is as follows:
\begin{equation}\label{eqn::gamma_variation}
    \sigma^2_{\gamma}(x,y) = \Bigg[\bigg(\frac{1}{\sum_{l}\sum_{s\in(x,y)} W_{ls}}\bigg)^2\sum_{l}\sum_{s\in(x,y)} (W_{ls})^2 \sigma^2_{e_{s}}\Bigg]/(1+K)^2,
\end{equation}
where $(1+K)$ is an additional multiplicative correction for the shear \citep{ellipticity_CFHTLens}. The sum is taken over all source galaxies $j$ that belong to a pixel, then sum over all LRG pairs, $l$. The scatter for a single lens-source pair $lj$ is $\sigma^2_{e_{s}} = \frac{\sigma^2_{\rm int}}{(w/w_{\rm max})}$, where $w$ is the weight assigned to each galaxy from $\textit{lensfit}$ algorithm. The intrinsic scatter, $\sigma_{\rm int}$, and the maximum weight, $w_{\rm max}$, are two constants taken as 0.28 and 16 respectively in CFHTLens survey. Then, this artificial scatter representing the shape noise is drawn from a Gaussian distribution with $N(0, \sigma^2_{\gamma}(x,y))$ and then added to the calculated shear value per pixel. This noise is propagated to the subsequent convergence calculations and finally to the subtraction between the two maps. After repeating the above procedure 1000 times, we estimate the average value and dispersion of the mass distribution measured from the enclosed box.
 
Following the above procedure, the measured enclosed mass of the filament in the rectangular box is $(4.03\pm0.89)\times10^{13} M_{\odot}$. Another empirical way of estimating the variance of the excess mass is to place boxes of the same dimensions on the resulting excess convergence map. Specifically, boxes are chosen with identical size of the filament region ($0.86\times0.30$ in the standardized coordinates), and they are placed in the region between $\pm$ 4 along the x axis and $\pm$ 4 along the y axis. A region centred at (0,0) between $\pm$ 1 along the x axis and $\pm$ 1 along the y axis is excluded in order to avoid the filament region. In this region, there are 210 such independent boxes, and the standard deviation of the excess mass is around $1.35 \times 10^{13} M_{\odot}$, which is comparable to but larger than the uncertainty estimated from the shape noise. However, the shape noise increases towards the edges of the map, compared to its value at the filament region, because there are fewer LRG-source pairs there.  We therefore adopt this value as a conservative estimate of our measurement uncertainty and propagate it into further analysis. A histogram of the masses in these boxes is shown in the bottom right panel of Figure \ref{lensing_plots}, where the blue dashed line shows the measured mass of the filament within the rectangular box in the centre.

The filament mass above is more than a factor two larger than found by \citet{Seth_filament_paper} from the same data. If instead we compare the mean convergence $\kappa$ within the box, our values are similar to those of \citet{Seth_filament_paper}, suggesting that there is no significant difference in the basic analyses. However, we note that \citet{Seth_filament_paper} quotes a critical surface mass density of 1640$M_{\odot}/\rm pc^2$, whereas we find 3471$M_{\odot}/\rm pc^2$ for LRG pairs at the same mean redshift. These two values can be reconciled if the one from \citet{Seth_filament_paper} is actually in units of solar masses per square \emph{comoving} pc. If this was erroneously multiplied by the area of the box in (\emph{proper} pc)$^2$, this would account for most of the discrepancy.
 
To measure the evolution of the filament mass with redshift, we perform the same mass measurement and uncertainty calculation for the LOWZ ($0.15< z_{\textrm{LRG}}<0.43$) and CMASS ($0.43<z_{\textrm{LRG}}<0.70$) subsamples separately. For the same filament box, we find $(4.86\pm2.54)\times10^{13} M_{\odot}$ for the LOWZ sample and $(3.63\pm1.84)\times10^{13} M_{\odot}$ for the CMASS sample. To combine these two measurements, we take a simple average of the LOWZ and CMASS values, which is shown as ``LOWZ+CMASS'' in Table \ref{table_summary_results}. This gives our final estimate of the enclosed mass of the filament for LOWZ and CMASS combined: $(4.25\pm1.57)\times10^{13} M_{\odot}$, which is a detection at the $\sim3\sigma$ significance level.

\begin{figure}
\includegraphics[width=\columnwidth]{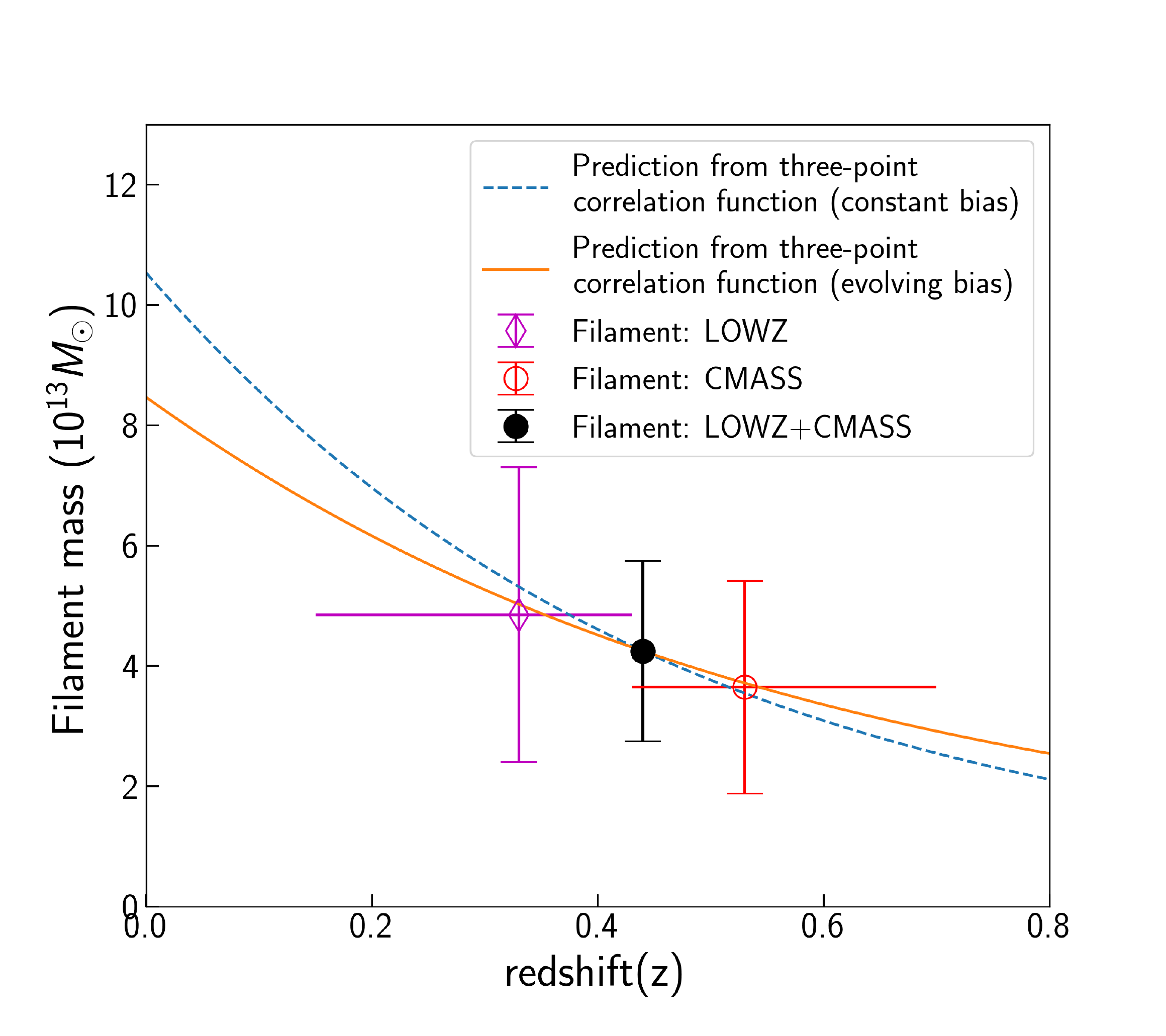}
   \caption{Filament mass as a function of redshift. The thin magenta diamond and red hollow circle with errorbars are the measurements from two independent samples LOWZ and CMASS. The solid black circle the average of the LOWZ and CMASS measurements. The blue dashed line shows the scaling of the filament mass expected from three-point correlation function with a constant bias for a cosmic filament with 2D $R_{\textrm{sep}}$= 6$h^{-1}$--10$h^{-1}$ Mpc between two $10^{13} M_{\odot}$ haloes. The orange solid line shows the prediction for an evolving bias model (see text for details).}
   \label{filament_mass_evolution}
\end{figure}

At face value, the observed mass of the filament appears to decrease with the increasing of redshift, although the difference is not statistically significant. Theoretically, the scaling of the filament mass with redshift can be predicted from evolution of the three-point correlation function which depends on the bispectrum \citep{Clampitt_2014, Seth_filament_paper, three_point_correlation_function_prediction}. Following \citet{Clampitt_2014}, the projected 3PCF among haloes at fixed locations $\Vec{x_1}, \Vec{x_2}$ and $\kappa$ at $\Vec{x_3}$ is given by
\begin{equation}\label{eqn::3PCF}
\Sigma_{m} (z) \equiv \Sigma_{\rm crit}\zeta_{gg\kappa} = \Sigma_{\rm crit}\langle \delta_{g}(\Vec{x_1})\delta_{g}(\Vec{x_2})\kappa(\Vec{x_3})\rangle \,.
\end{equation}
where $\zeta_{gg\kappa}$ can be calculated from the bispectrum \citep{Clampitt_2014}. The redshift evolution of $\Sigma_{m} (z)$, to leading order in perturbation theory, scales as
\begin{equation}
    \Sigma_{m} (z) = \left(\frac{b(z)}{b(z=0)}\right)^2 \left(\frac{D(z)}{D(z=0)}\right)^4 \Sigma_{m}(z=0) \,,
\end{equation}
where $b$ is the linear bias of LRGs and  $D(z)$ is the linear growth factor. The redshift scaling is shown in Figure \ref{filament_mass_evolution}, after fixing the normalisation to agree with the mean filament mass at the mean filament redshift. For the evolution of galaxy bias, we show two possibilities. One uses a constant galaxy bias, which we set to be $b=2.04$, the average of $b = 2.08$ for LOWZ case and $b = 2.01$ for CMASS case \citep{LRG_halo_bias_LOWZ_CMASS_separate}. Although the uncertainties are large, the data are consistent with the predicted redshift scaling. The other possibility is an evolving bias model $b-1 = (b_{0}-1)/D(z)$ \citep{evolved_bias_model}, where $b_{0}$ is the linear bias extrapolated to present day.

Recently, \citet{new_filament_paper} claimed a 3.3 $\sigma$ level detection of an anisotropic shear signal from the filament using only LOWZ LRG pairs, but combining shapes from KiDs+VIKING-450 Survey (KV450), the Red Cluster Sequence Lensing Survey (RCSLenS) and the CFHTLenS. In their study, the selection criteria for filament candidates were slightly different from ours. They select all pairs with 2D projected separation $3h^{-1} \textrm{Mpc} \leqslant R_{\textrm{sep}} \leqslant 5 h^{-1} \textrm{Mpc}$ (instead of $6h^{-1} \textrm{Mpc} \leqslant R_{\textrm{sep}} \leqslant 10 h^{-1} \textrm{Mpc}$ as adopted by this study). Using the same method as presented above, we re-run the analysis to measure the mass of the filament from our data using their LRG separation criteria. We find $M_{3-5} = (4.35\pm1.35)\times10^{13} M_{\odot}$, where the uncertainty comes from shape noise only. The mass measured from their study is $M_{\rm Xia+} = (6.7\pm3.1)\times10^{13} M_{\odot}$, but this mass is obtained by fitting a power law density profile to the filaments. \citet{Intercluster_filaments_in_Universe} studied filaments in N-body simulations and found that, in cylindrical coordinates aligned with the filament axis, the density profile has a core and then falls like $r^{-2}$ at large radii.  However, \citet{Mead_filament_paper} modelled the \emph{projected} convergence map of the filaments with the same functional form:
\begin{equation}\label{qgn::colberg_fil_density_profile}
    \kappa(r) = \frac{\kappa_{0}}{1+\left(\frac{r}{r_{c,2d}}\right)^2},
\end{equation}
and \citet{new_filament_paper} used this model in their work. Their filament mass is defined by integrating the profile in the direction perpendicular to the filament from 0 to infinity, and along the filament over a range of length $R_{\rm sep}$ yielding a mass of $\pi r_{c,2d} R_{\rm sep} \mathcal{F}_{c}$. In our analysis, the integration is conducted in a projected rectangle, where the height of the rectangle is $0.30R_{\rm sep}$ and the length is $0.86R_{\rm sep}$. We use their best-fit parameters of the density profile (presented in column ``All'' in their Table 3) and calculate the expected mass ratio between their measurement and ours as
\begin{equation}\label{eqn::mass_factor_comparison}
   \frac{M_{3-5}}{M_{\rm Xia+}} = \frac{2\times0.86\int_{0}^{0.15R_{\rm sep}} \frac{\mathcal{F}_{c}}{1+(\frac{r}{r_{c,2d}})^2}dr}{2\times \int_{0}^{\infty} \frac{\mathcal{F}_{c}}{1+(\frac{r}{r_{c,2d}})^2}dr} = 0.54,
\end{equation}
where the value of $R_{\rm sep}$ is adopted as $4h^{-1} \textrm{Mpc}$. Thus given the differences in definition, we expect our mass to be 54\% of theirs and we measure a ratio of 0.65, and so we conclude these are consistent.
 
\section{Maps of light and stellar mass in the filament}\label{sec:: light_map}
 
In this section, we turn to the luminosity and stellar mass of the filament, which is the main focus of this work. Our method for measuring  the light and stellar mass maps are similar to the construction of lensing map: we will stack galaxy luminosity for physical and non-physical pairs and subtract them to isolate the excess light associated with the filament. However, there are some subtle differences between these two situations. In Section \ref{ssec::light_pair_selection} and \ref{ssec::data_selection_source_galaxies_light}, we discuss the LRG pair selection and galaxies used in the construction of the light map. In Section \ref{sssec::observed_light_map}, we demonstrate our 2D projected luminosity and stellar mass density map. In Section \ref{sssec::total_luminosity}, we present our method of measuring the total light content of filaments and some discussions on the results.
 
\subsection{LRG pair selection for the light map}\label{ssec::light_pair_selection}
 
In addition to the $z_{\rm sep}$ and 2D $R_{\rm sep}$ selection cuts mentioned in Section \ref{ssec::data_selection_LRG_pairs}, to enhance the actual filament signal and suppress the background noise, for the light and stellar mass maps, we select all the LRGs within region $120^{\circ} \leqslant \textrm{RA} \leqslant 230^{\circ}$, $10^{\circ} \leqslant \textrm{Dec} \leqslant 60^{\circ}$, rather than restricting ourselves only to the CFHTLenS area. From the whole BOSS redshift coverage, there are 448,314 LRG pairs with <$z_{\textrm{pair}}$> = 0.44 and <$R_{\textrm{sep}}$> = 8.04 $h^{-1}$ Mpc. For the LOWZ sample, we find 50,917 pairs with <$z_{\textrm{pair}}$> = 0.33 and <$R_{\textrm{sep}}$> = 8.03 $h^{-1}$ Mpc. The CMASS sample has 390,748 pairs with <$z_{\textrm{pair}}$> = 0.53 and <$R_{\textrm{sep}}$> = 8.04 $h^{-1}$ Mpc. 

The lensing map was constructed by taking an ensemble average of the galaxy shapes over all LRG pairs. Notice that the number of galaxies entering into the average does not bias the result, although it does affect the noise. Consequently, the weak lensing mass map is not affected by incompleteness in the source catalogue. In contrast, for the light map, it is necessary to count precisely the number of galaxies per LRG pair located in each area of the map. Furthermore this counting is done in rescaled-coordinates, and a given unit area will subtend a larger area on the sky for an LRG-pair at lower redshift or if $R_{\rm sep}$ is larger. This means that in order that the physical and non-physical light maps are comparable, the redshift and 2D $R_{\textrm{sep}}$ distribution of two LRG-pair catalogues need to be as close as possible. In general, the non-physical pairs are randomly distributed, so the count of the non-physical pairs should increase linearly as a function of $R$ separation, but this is not the case for the physical LRG pairs (which are clustered). Therefore, there is a higher fraction of non-physical pairs at large $R$ separation bins than those of physical pairs, which would lead to differences in the mean galaxy number density between the physical and non-physical projected maps. To correct this, we apply a one-to-one match between physical and non-physical pairs based on their redshift and 2D projected separation by finding all non-physical pairs, where the difference $\Delta$ between the physical and non-physical pairs,  limited to $\Delta z$ < 0.01 and $\Delta R_{\rm sep}$ < 0.2, and only keeping the one with the smallest $\Delta R_{\rm sep}$.

In this way, we generate a catalogue of non-physical pairs which has approximately the same redshift and $R_{\textrm{sep}}$ distributions as physical pairs, ensuring that the mean foreground/background level projected in the same filament box is comparable. 
 
\subsection{Photometric galaxy catalogue for the light map}\label{ssec::data_selection_source_galaxies_light}

To produce a stacked light map, we use galaxies from the SDSS catalogue. Specifically, the photometric catalogue was constructed by cross-matching GalaxyTag and PhotoZ in SDSS DR14 \citep{photoz_SDSS}. Galaxies that are marked with photometry flag $\textsc{CLEAN = 1}$, bestFitTemplateID > 0 and $z_{\rm photo}>0$ are selected. The first criterion ensures the reliability of the photometric measurements, and the last two criteria are used to remove bad measurements. We include all galaxies with $r$-band cmodel magnitude brighter than 21.5. Note that, for LRG pairs located close to the boundaries on the sky of the photometric catalogue, photometric galaxies are only available on one side of the pair but lacking on the other side. To avoid this edge effect, the sky coverage of the source galaxies is chosen to be slightly larger ($\sim$ 8 degree) than the coverage of LRG spectroscopic catalogue. This yields a sample of 29,703,867 galaxies in the $r$ band.

In order to investigate the colour of the galaxies in the filament, we also conduct the same analysis in the $g$-band, and the selection criteria for the photometric data are identical except for the choice of apparent magnitude limit which is set to 22.5\footnote{The choice of r and g band magnitude cut are more or less tentative. However, while computing the total luminosity, our final results are not sensitive to the choice of magnitude cut.}. These magnitude limits corresponds to approximately 50$\%$ completeness for each band.

To compute the luminosity of a galaxy, we use the SDSS $\textsc{Photoz}$ table which contains $z_{\textrm{photo}}$, spectral type, and $K$-corrections for all primary objects flagged as galaxies. The photometric redshifts are determined using the ``empirical method'' as discussed in \citet{photoz_SDSS}. The $K$-corrections are based on the best-fitting spectral template from \citet{An_atlas_of_composite_spectra} for each galaxy at its photometric redshift.

The luminosity is then
\begin{equation}\label{eqn::lum}
L_{^{0.1}n}/L_{\odot, ^{0.1}n} = 10^{(M_{\odot, ^{0.1}n} - M_{^{0.1}n})/2.5}\, ,
\end{equation}
where
\begin{equation}\label{eqn::abs_mag_cal}
M_{^{0.1}n} = m_{n} - \textrm{DM}(z_{\textrm{photo}}, \Omega_m = 0.3, \Omega_{\Lambda} = 0.7, h = 0.7) - K_{^{0.1}{n}}\, ,
\end{equation}
DM $= 5\log_{10} (D_{\rm lum}/10\,\textrm{pc})$ and where $n$ refers either to the $^{0.1}r$- or the $^{0.1}g$-band. We adopt these shifted bands to make the later comparison with SDSS literature results simpler.

Having calculated the luminosity, the stellar mass of a galaxy can be obtained from the stellar-mass-to-light ratio, $M_{\rm stellar}/L_{^{0.1}r}$. As mentioned previously, for the galaxies in the $\textsc{Photoz}$ Table, each galaxy is matched with a best-fitting template from which physical parameters, such as $M_{\rm stellar}/L_{^{0.1}r}$, are also estimated \citep{An_atlas_of_composite_spectra}. 

There are 53,453 galaxies (approximately 0.2\% of the total sample) with calculated luminosity or stellar mass greater than $1\times10^{12} L_{\odot}$ or $1\times10^{12} M_{\odot}$ for both $r$ and $g$ band. We assume that most of these are due to catastrophic failures of the photometric redshifts and discard them from the samples.

\subsection{Light and stellar mass maps}\label{sssec::observed_light_map}

\begin{figure*}
    \begin{minipage}[b]{0.5\textwidth}
        \centering
        \includegraphics[width=1.05\linewidth]{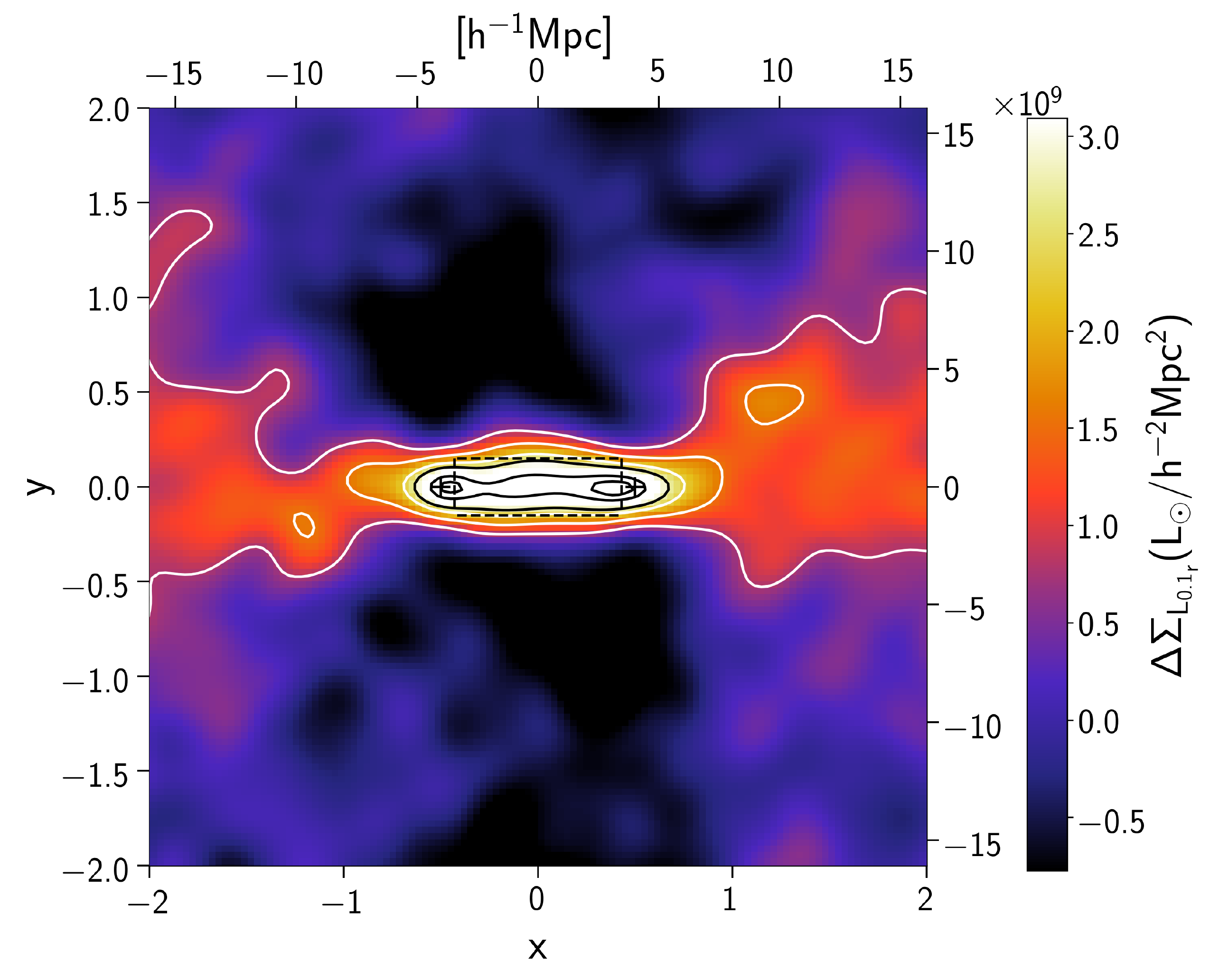}
    \end{minipage}\begin{minipage}[b]{0.5\textwidth}
        \centering
        \includegraphics[width=0.9\linewidth]{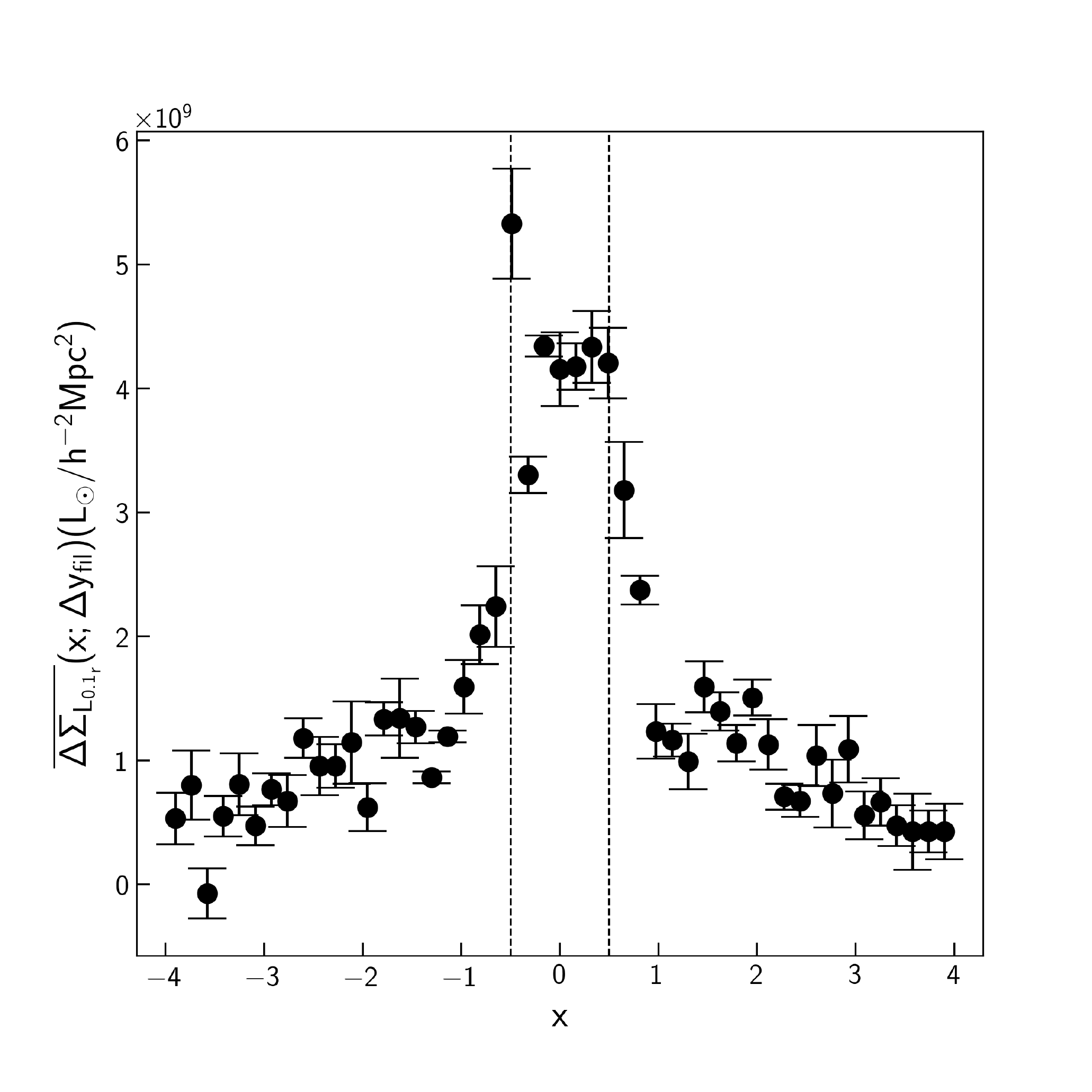}
    \end{minipage}\\
    \begin{minipage}[b]{0.5\textwidth}
        \centering
        \includegraphics[width=1.05\linewidth]{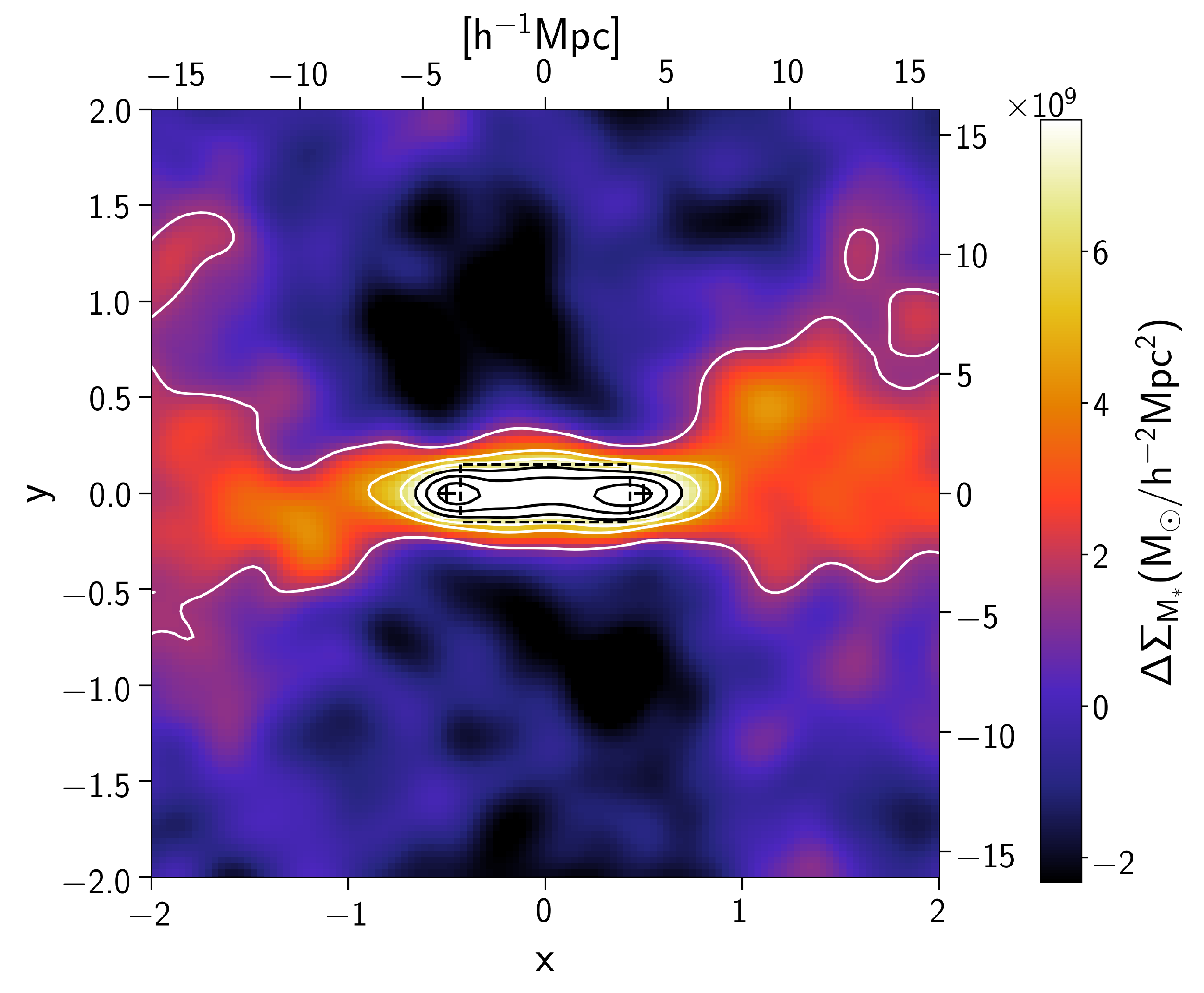}
    \end{minipage}\begin{minipage}[b]{0.5\textwidth}
        \centering
        \includegraphics[width=0.9\linewidth]{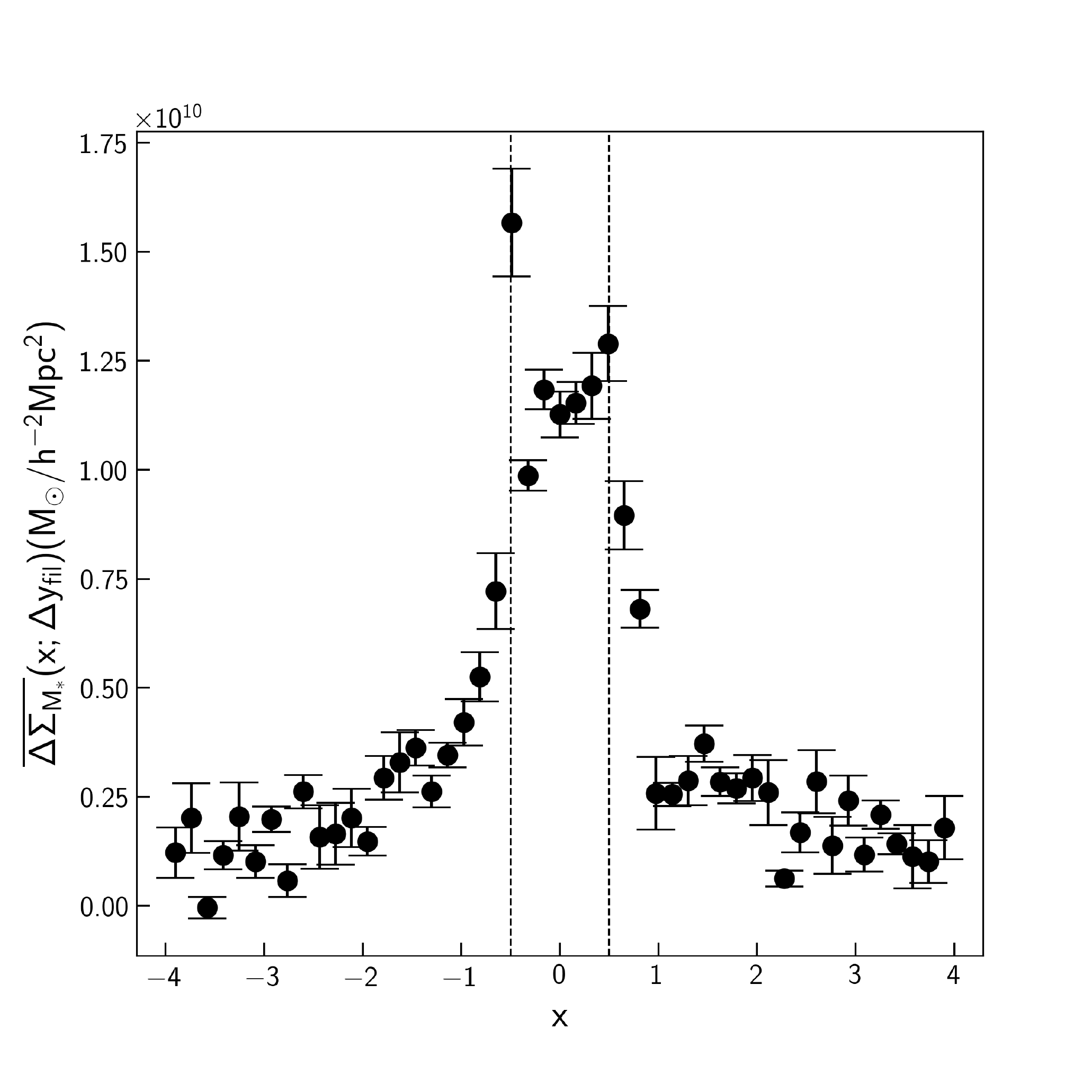}
    \end{minipage}%
\caption{The excess luminosity (top row) and stellar mass density (bottom row) maps per LRG pair obtained from the whole sample of $\sim$ 450,000 LRG pairs. The units are solar luminosity per $h^{-2}\textrm{Mpc}^2$ for luminosity density and solar mass per $h^{-2}\textrm{Mpc}^2$ for stellar mass density. One unit in the rescaled coordinate system corresponds to approximately 8 Mpc $h^{-1}$ proper size. Left panel: The projected 2D excess luminosity/stellar mass density map obtained by subtracting the non-physical pair map from the map of physical pairs.  A Gaussian filter of smoothing scale 0.90 $h^{-1}\rm Mpc$ (0.10 in units of $x,y$) has been applied to the image to suppress small-scale noise.  The locations of LRG haloes in the standardized coordinate system are marked with two ``+'' symbols. There is significant excess light at and between the two LRG pairs. Right panel: The profile of excess luminosity/stellar mass density along the $x$ axis. The locations of LRG haloes are indicated by the two black dashed lines. At each point of $x$, the values of $\Delta{\Sigma_{L}}$ (without smoothing) are averaged along the fiducial height $\Delta{y}$ = 2.45 Mpc $h^{-1}$ (0.3 in units of the rescaled coordinate system).}
\label{lum_stellar_obs_map_all}
\end{figure*}

We then stack the luminosities (or stellar masses) of SDSS galaxies around $\sim$ 450,000 LRG pairs, and subtract non-physical pairs from the physical pairs to produce the observed excess luminosity and stellar mass density per LRG pair shown in Figure \ref{lum_stellar_obs_map_all}. Here, we only show the results obtained from the whole sample as this is less noisy. The left hand panels show the 2D projected map, where the rectangular box delineates the size of projected filament region. The values of $\Delta{\Sigma_{L}}$ and $\Delta{\Sigma_{M_{\rm stellar}}}$ computed from the central filament box are $(0.68\pm0.04) \times 10^{11}L_{\odot, ^{0.1}r}$ and $(1.91\pm0.09) \times 10^{11} M_{\odot}$, where the errorbars are evaluated from the covariance matrix described below. To further show the significance of our measurements, on the right panels of both plots, we plot the averaged excess luminosity and stellar mass density as a function of the $x$ coordinate: for each $x$, we averaged $\Delta{\Sigma_{L}}$($\Delta{\Sigma_{M_{\rm stellar}}}$) along the $y$-axis. There is a highly significant detection of light and stellar mass in the filament region.

These maps show the ``dumbbell'' structure expected from the three-point correlation function, as calculated in \citet{Clampitt_2014} and \citet{Seth_filament_paper}: the excess light peaks at the locations of the LRG haloes, with a filament connecting them. Furthermore, the filament extends beyond the two LRG haloes, albeit at lower density. These maps are significantly less noisy than the WL mass maps due primarily to the much larger number of LRG pairs used. 

\begin{figure*}
    \begin{minipage}[b]{0.5\textwidth}
        \centering
        \includegraphics[width=\linewidth]{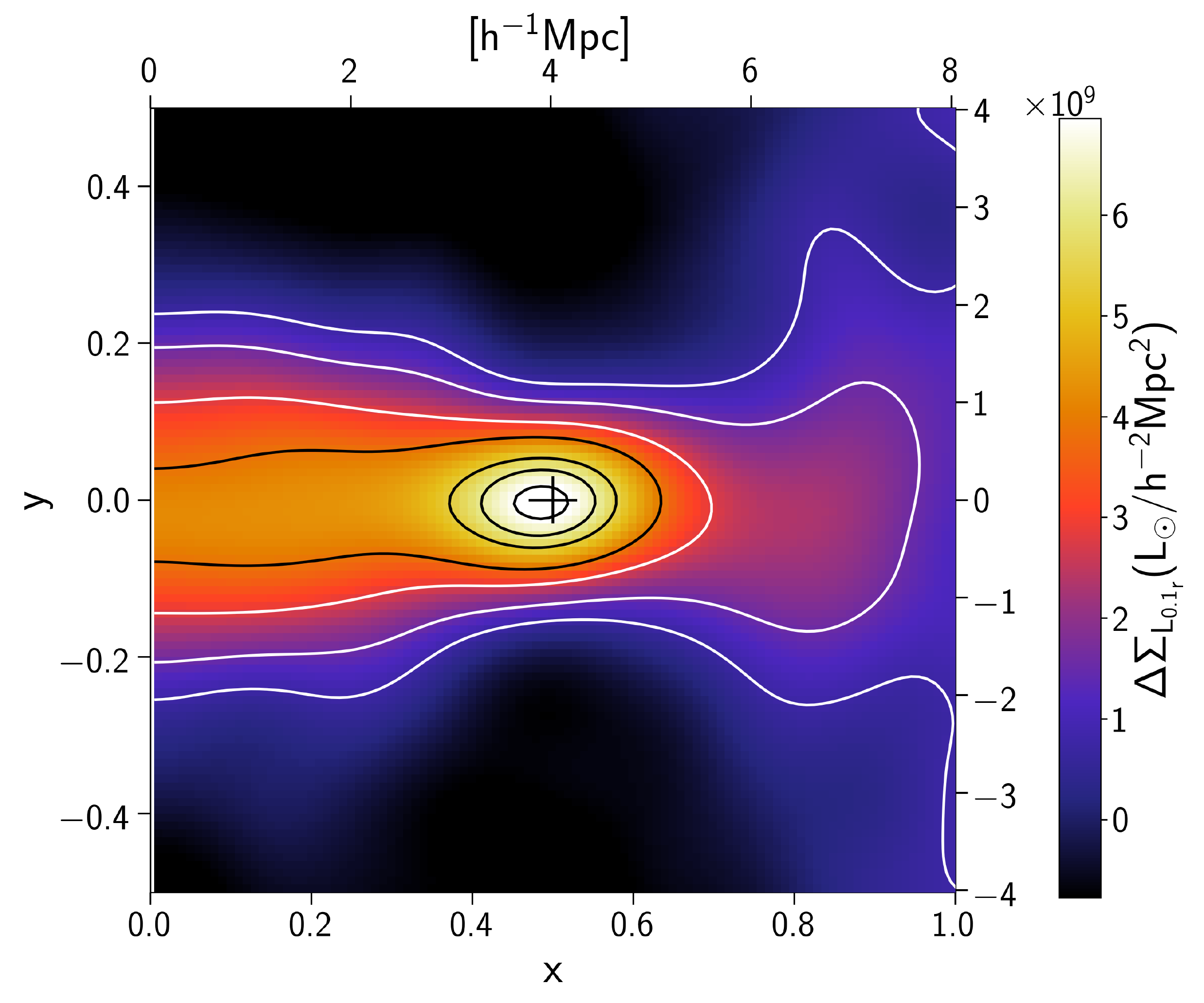}
    \end{minipage}\begin{minipage}[b]{0.5\textwidth}
        \centering
        \includegraphics[width=\linewidth]{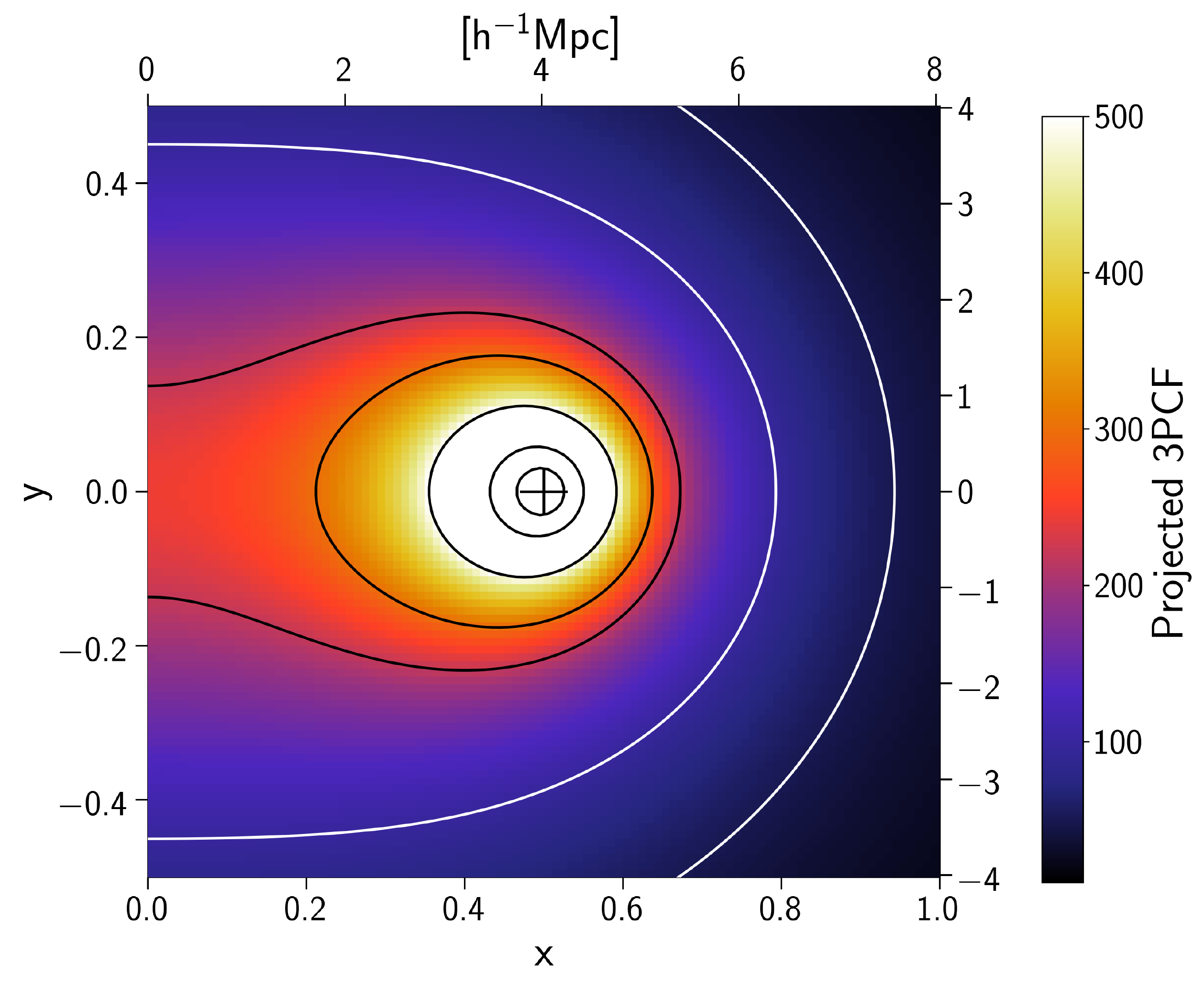}
    \end{minipage}%
\caption{Left panel: The excess luminosity density map with higher resolution, zooming in on the region around the light peak. To reduce the noise, we mirror the left part of the light map about the y-axis and average it with the right part. A Gaussian filter of smoothing scale 0.50 $h^{-1}\rm Mpc$ (0.065 in units of $x,y$) has been applied to the image for illustration. The location of LRG is marked with ``+'' symbol.  Right panel: Contour plot of the projected three-point
correlation function around the excess peak.}
\label{filament_light_zoom_in}
\end{figure*}

To better study the light distribution around the location of LRG, we zoom in on the region around the light peak, using finer pixels in the map. We also mirror the left part of the light map about the $y$-axis and average it with the right part in order to reduce the noise. The resulting luminosity density map is shown in Figure \ref{filament_light_zoom_in}. From this zoomed-in map, one can see that light distribution around the peak is lopsided with its centroid shifted in the direction of the other LRG. This feature is similar to distribution of satellites around Milky-Way mass halos reported in \citet{lopsided_distribution_satellite}. 

In fact, the lopsidedness is expected from the three-point correlation function (3PCF). For simplicity, here, we simply approximate the projected 3PC using the power-law model: $\zeta = C [(r_{L\kappa}/r_{0})(r_{R\kappa}/r_{0})]^{-\gamma}$ \citep{3PCF_paper}, valid when $r/r_0 \ll 1$, with two points are held at fixed positions (the location of LRG haloes) and the third point is located at the position at which we evaluate $\zeta$. Specifically, $r_{L\kappa}$ ($r_{R\kappa}$, respectively) is the distance between the location of the left (right) LRG halo and the third point on the map. Here $\gamma$ is taken to be 2 and the correlation length $r_{0}$ is $10h^{-1} \rm Mpc$. We fix the value of normalisation $C$ to 1 because our main purpose here is to compare the predicted structure and shape to the observation. The resulting contour plot of the projected three-point correlation function is shown on the right panel of Figure \ref{filament_light_zoom_in}. Interestingly, three-point correlation function indeed predicts a lopsided distribution around the excess peak. However, the predicted filament seems to have a thicker structure in the $y$-direction than the observed one. These results are consistent with \citet{Seth_filament_paper} (their Figure 7), where they performed a more precise prediction to the 3PCF using linear perturbation theory and bispectrum (see Equation \ref{eqn::3PCF}).

\begin{figure*}
    \begin{minipage}[b]{0.5\textwidth}
        \centering
        \includegraphics[width=1.1\linewidth]{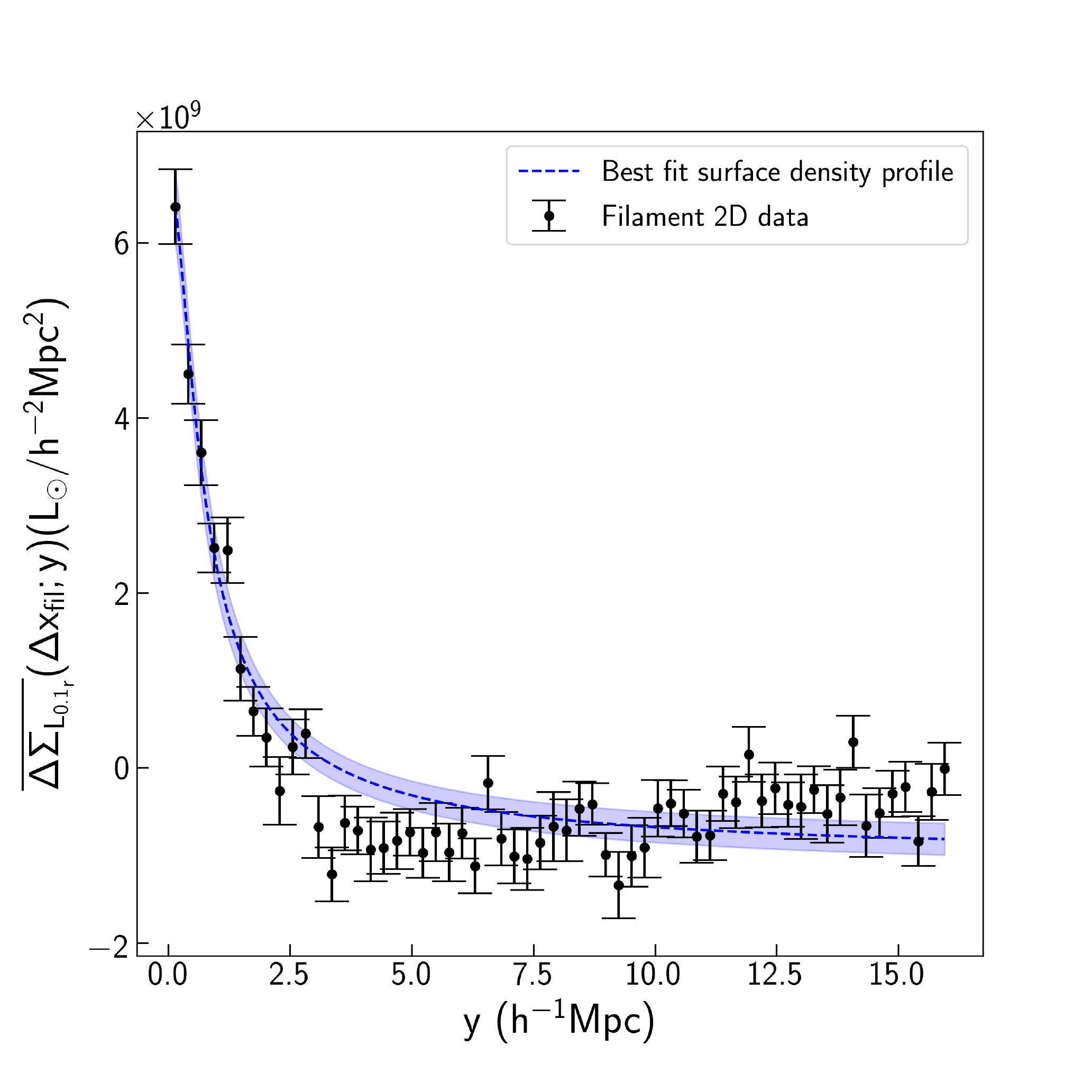}
    \end{minipage}\begin{minipage}[b]{0.5\textwidth}
        \centering
        \includegraphics[width=1.1\linewidth]{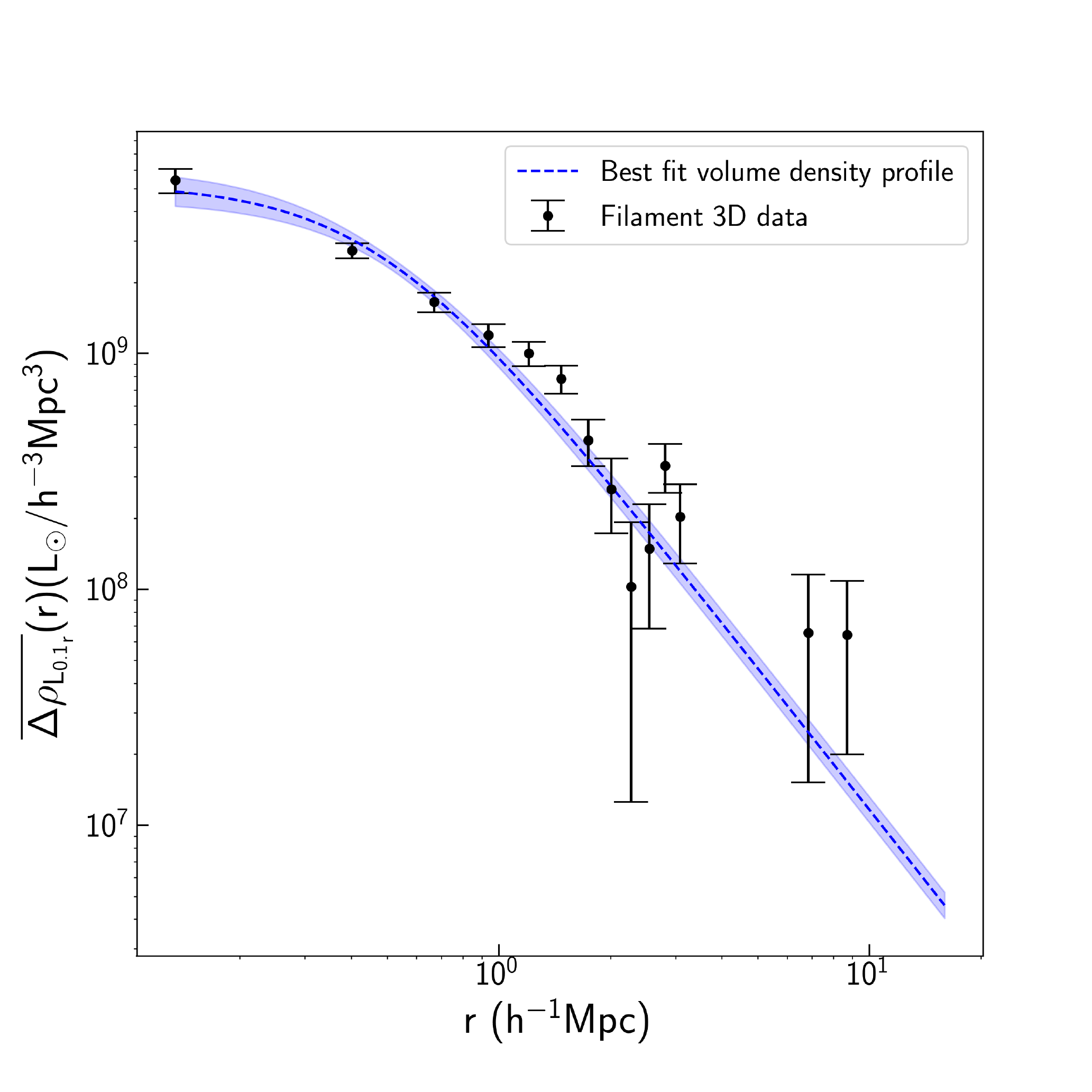}
    \end{minipage}%
\caption{Left panel: the profile of excess luminosity surface density along the $y$ axis. At each value of $y$, the values of $\Delta{\Sigma_{L}}$ (without smoothing) are averaged along the fiducial filament length $\Delta{x}$ = 7.0 Mpc $h^{-1}$ (0.86 in units of the rescaled coordinate system). Right panel: the profile of excess luminosity volume density as a function of distance perpendicular to the cylinder axis, where the volume density has been obtained by Abel-transforming the observed 2D data. The best fit and its standard deviation are over-plotted on the same figure.}
\label{lum_map_all_excess_along_y}
\end{figure*}

It is interesting to compare our observations to the 3D cylindrical density profile suggested by \citet{Intercluster_filaments_in_Universe}
\begin{equation}\label{eqn::fil_3D_profile_along_y}
    \rho(r) = \frac{\rho_0}{1+\left(\frac{r}{r_c}\right)^2}\, ,
\end{equation}
where $r$ is perpendicular to the cylinder axis, $r_c$ is a core radius and $\rho_0$ is the central density along the axis.
To fit the data, we first compute the projected 2D density profile of Equation \ref{eqn::fil_3D_profile_along_y} using the Abel transform, which is given by:
\begin{equation}\label{eqn::fil_2D_profile_along_y}
 \Sigma(y) = \frac{\rho_0\pi{r_c}^2}{\sqrt{{r_c}^2 + y^2}} - \textrm{constant},   
\end{equation}
where a constant is added to the 2D profile in order to account for the negative values that appear in the observations at large $y$. Note that the addition of a constant to the 2D profile does not change the original form of 3D profile from the Abel inversion. The averaged excess luminosity surface density as a function of the $y$ coordinate is shown on the left panel of Figure \ref{lum_map_all_excess_along_y}: $\Delta{\Sigma_{L}}$ is averaged along the $x$-axis for each $y$. The surface density is decreasing and becomes roughly constant at large $y$. For each data point, the plotted error bar is taken from the diagonal elements in the covariance matrix of $\Delta{\Sigma_{L}}$ measurements, while in the fitting procedure, the full covariance matrix is used. Data points are fitted using Equation \ref{eqn::fil_2D_profile_along_y}. The best-fit line and its standard deviation are over-plotted on the same figure, where the best-fit parameters are $\rho_0 = (5.25\pm0.90)\times10^9 L_{\odot}/h^{-3}\rm Mpc^3$, $r_c = (0.47\pm0.07) h^{-1}\rm Mpc$ and const = $(1.04\pm0.20)\times10^9 L_{\odot}/h^{-2}\rm Mpc^2$. Note that $\rho_0$ and $r_c$ are highly anti-correlated (with correlation coefficient $\sim -0.93$). The right panel shows the profile of excess luminosity volume density along the cylindrical $r$ axis in log-log scale. The black solid points are obtained by Abel-transforming the observed 2D data. To estimate the uncertainties of the transformed data points, we draw random samples from a multi-normal distribution using the full covariance matrix of $\Delta{\Sigma_{L}}$ measurements. The above procedure is repeated 1000 times, and the resulting average values are shown as black solid points, with black error bars representing the standard deviations.

Assuming a constant mass-to-light ratio, we can further compute the best-fit 3D mass by integrating the 3D luminosity profile. The integral of the profile given by Equation \ref{eqn::fil_3D_profile_along_y} does not converge at large radii so it is necessary to specify a cylindrical radius within which the luminosity and mass are well defined. We adopt a cylindrical radius of 1.2$h^{-1} \rm Mpc$, which matches the height of the projected rectangular box discussed previously. The resulting excess 3D mass is $M_{\rm cyl, 3D} = (3.32 \pm 1.22) \times 10^{13} M_{\odot}$. Then, assuming that the filament is a cylinder of length\footnote{Note that this length is slightly longer than the 2D projected length (0.86$\times8 h^{-1} \rm Mpc$) in order to account for the tilt of the cylinder axis with respect to the plane of the sky.} 8 $h^{-1} \rm Mpc$ and radius 1.2$h^{-1} \rm Mpc$, the corresponding excess density within the cylinder is $\bar{\delta}_{\rm cyl, 3D} = M_{\rm cyl, 3D}/M_{\rm b} = 2.58\pm0.98$, where $M_{\rm b}$ is the background mass estimated at $z = 0.44$ within the cylinder. 

We can compare our result with the measurement obtained by \citet{de_graaf_fil}. They found a 1.9 $\sigma$ detection from CMB lensing with $\bar{\kappa} = (0.6\pm0.3)\times10^{-3}$ from the stacked filament between CMASS LRG galaxy pairs. They proposed a cylindrical model of filament in which the total matter distribution follows a two-dimensional Gaussian profile. Using their model and best-fit parameters, we integrate their profile in the same filament cylinder mentioned above (with length $= 8 h^{-1} \rm Mpc$ and radius $=1.2h^{-1} \rm Mpc$). The resulting excess density within the cylinder is $\bar{\delta}_{\rm cyl, 3D} = 2.12\pm1.12$, which is consistent with our estimate.

Using the Planck tSZ map, \citet{warm_hot_gas_filament} detected a level of 5.3 $\sigma$ tSZ residual filament signal with $\bar{y} = (1.31\pm0.25)\times10^{-8}$ between SDSS DR12 LRG galaxy pairs ($z$ peaked at $\sim 0.35$). In their study, they found that the best-fit profile of cylindrical filament model follows $r/r_{c}$ at large radii, which is different from the profile used in our study. We also tried fitting our data to the best-fit profile as adopted by \citet{warm_hot_gas_filament}. However, we found that Equation \ref{eqn::fil_3D_profile_along_y} fits our measurements better. To make a fair comparison, we calculate the predicted mean tSZ residual signal by integrating our best-fit model (Equation \ref{eqn::fil_3D_profile_along_y}) within their filament cylinder (with length $= 6.4 h^{-1} \rm Mpc$ and radius $=0.8h^{-1} \rm Mpc$). Then, using their observational mean tSZ signal, we can solve for the electron temperature $T_{e}$. Our resulting $T_{e}$ is found to be $(4.66\pm2.05)\times10^6 \rm K$. This is lower than the $(8.2\pm 0.6)\times10^{6} \rm K$ found in their simulations. However, our estimate is still consistent with their result within $2\sigma$ level given that the chosen best-fit density profiles and central over-density values are different in two studies.

\subsection{Total luminosity and stellar mass in the filaments}\label{sssec::total_luminosity}

The simple analysis discussed in the previous section, while useful for highlighting the spatial structure of the filament, has disadvantages for estimating the total light in the filament, due both to the flux limit of the SDSS $r$-band data, as well possible outliers in the photometric redshifts. We now discuss an alternative method that is less sensitive to these issues.

In order to measure the total light, within the 2D projected filament box, we fit a Schechter function to the excess galaxy counts (per LRG pair) as a function of apparent magnitude. This allows us to compute the total luminosity by integrating over the best-fitting Schechter function. Unlike the luminosity/stellar mass map,  excess galaxy number counts do not depend on photometric redshifts. We follow the same selection criteria and corrections for the construction of physical and non-physical LRG pairs catalogue. The number of galaxies is then divided into 100 equal-sized apparent magnitude bins from 15 to 21.5 (15-22.5 for $^{0.1}g$ band), and subtraction gives us the observed excess number as a function of magnitude.. Then the best-fit parameterization of the Schechter function can be obtained by comparing the observation with forward-modelled predicted counts. In each apparent magnitude bin, the predicted galaxy number for a given LRG pair can be computed via a Schechter function \citep{schechter_function_cite} by 
\begin{equation}\label{eqn::lum_function}
N(m, z) = \phi_{\star}\int_{L_{\textrm{min}}(z)}^{L_{\textrm{max}}(z)} \Big( \frac{L}{L_{\star}} \Big)^{\alpha} e^{-L/L_{\star}} \frac{dL}{L_{\star}} \,,
\end{equation}
where $\alpha$ is the slope of the luminosity function and $L_{\star}$ is the characteristic galaxy luminosity. The upper and lower luminosity limits in the integral can be computed with Equation \eqref{eqn::abs_mag_cal} based on the distance modulus and $K$-corrections of the filament galaxies. Because the spectroscopic redshifts of the LRG pairs are known, we assume the redshifts  of the galaxies in the filaments are identical to that of their host LRG pair. For the $K$-correction, we make a histogram of the excess number of galaxies in the filamentary box as a function of $K$-correction, then the weighted average of the distribution is our final choice of $K$-correction. Finally, we average over the redshifts of the LRG pairs.

Note that the normalization factor defined here is different from the usual pre-factor in the Schechter function. To make the normalization factor more physical, we further define $n_{\star}$ as the ratio between the total integrated luminosity and $L_{\star}$:
\begin{equation}\label{eqn::n_star_equation}
    n_{\star} = \frac{L_{\rm tot}}{L_{\star}},
\end{equation}
where $L_{\rm tot}$ is the total luminosity and index from the MCMC fits (as discussed below). $L_{\star}$ is calculated by using the best-fit $M_{\star}$ from MCMC.

Comparing the model to the observation, there are three different parameters to be determined: the normalization factor, $\phi_{\star}$, the slope, $\alpha$, and the characteristic galaxy absolute magnitude, $M_{\star}$. To find the best-fitting set of parameters, we define the following $\chi^2$:
\begin{equation}\label{eqn::chi_square_lum_fit}
\chi^2(\phi_{\star}, \alpha, M_{\star}) = [N_{\textrm{obs}}(m) - N_{\textrm{mod}}(m)]^TC^{-1}[N_{\textrm{obs}}(m) - N_{\textrm{mod}}(m)],
\end{equation}
where $N_{\textrm{obs}}(m)$ is the actual observation and $N_{\textrm{mod}}(m)$ is the expected number per pair in each apparent magnitude bin integrated using Equation \eqref{eqn::lum_function} and $C$ is the sample covariance matrix. To compute the covariance matrix, for each LRG pair, we record the excess number of galaxies as a function of apparent magnitude in the central rectangular region. There are $N_{\rm pair}$ observations in total, and these observations are treated as input of the \textit{numpy.cov} function, which is used for calculating the covariance matrix. We then use the best-fitting parameters to compute the total light by integrating the Schechter function over luminosity from 0 to $\infty$.

\begin{figure*}
    \begin{minipage}[b]{0.5\textwidth}
        \centering
        \includegraphics[width=1.1\linewidth]{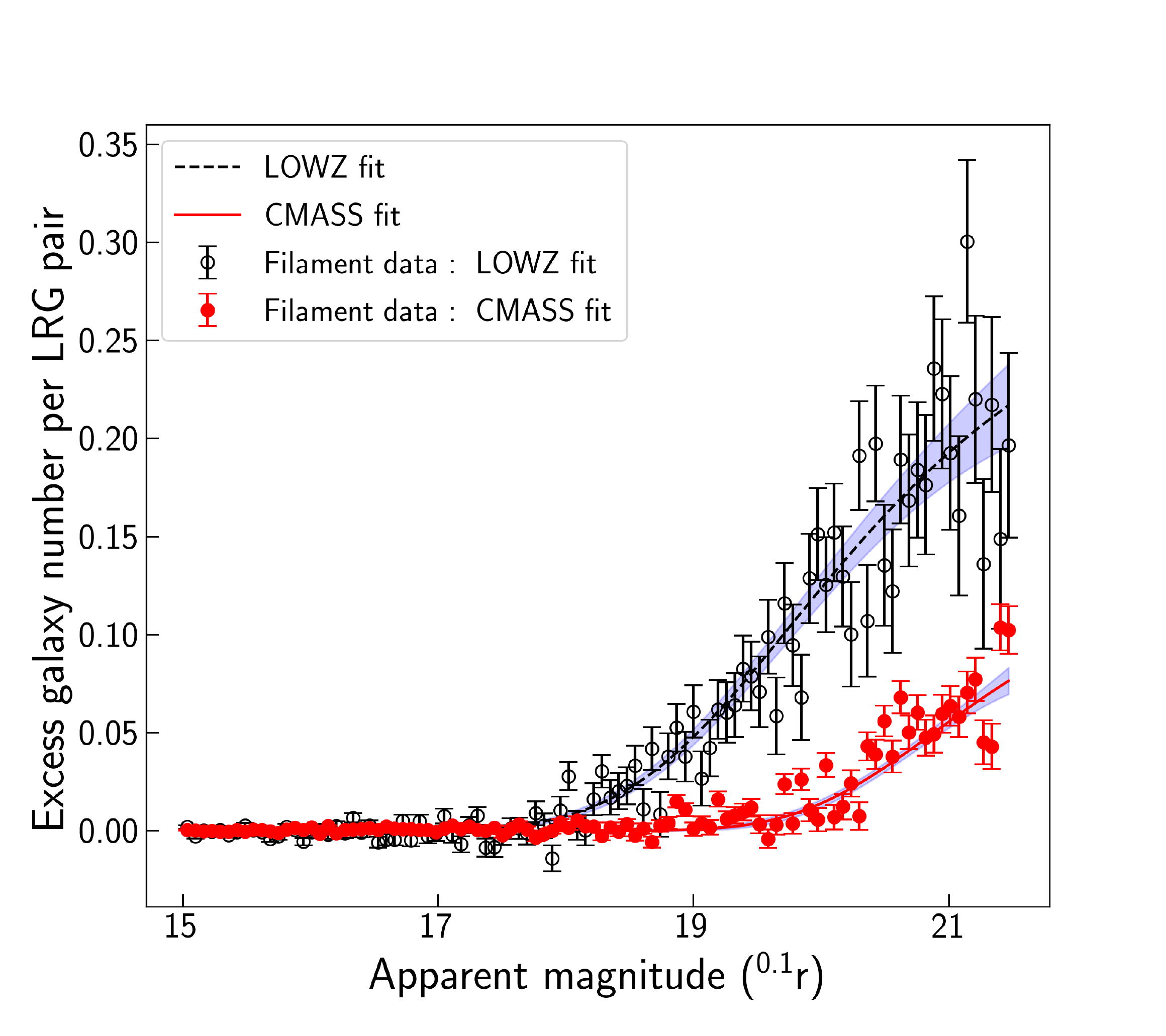}
    \end{minipage}\begin{minipage}[b]{0.5\textwidth}
        \centering
        \includegraphics[width=1.1\linewidth]{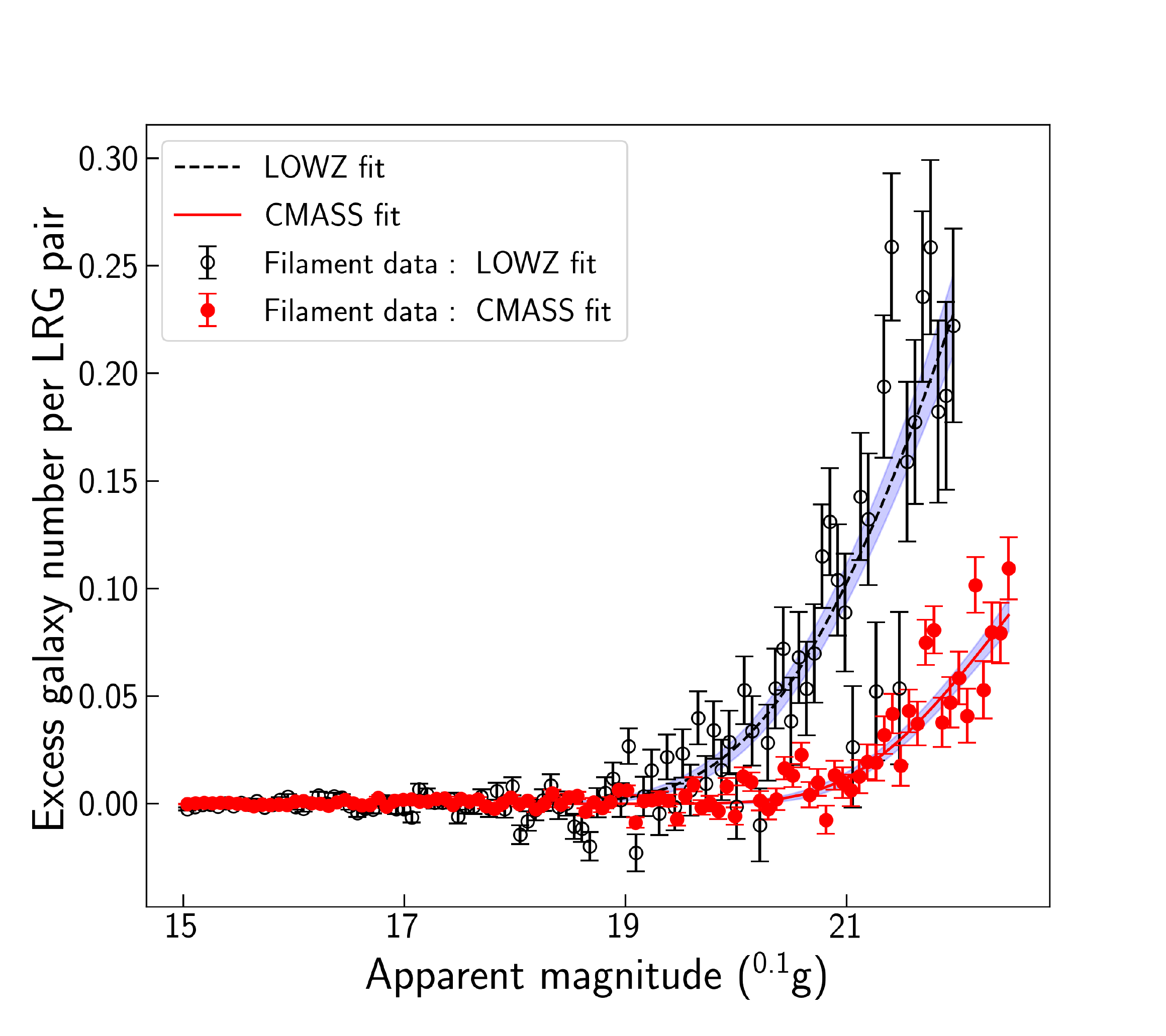}
    \end{minipage}%
\caption{The observed excess number of galaxy per LRG pair as a function of apparent magnitude. The best-fit Schechter function for LOWZ and CMASS samples for different bands are over-plotted on the same figure. The best-fit line and its standard deviation are computed from 50 MCMC samples. The left panel shows the data and the fit in $^{0.1}r$ band, and the right panel shows the result in $^{0.1}g$ band. }
\label{schechter_function_plot}
\end{figure*}

To obtain the best-fit parameterizations of the Schechter function, we use the Markov chain Monte Carlo (MCMC) method. More specifically, we use the online Python package \textsc{emcee} \citep{emcee_paper} with the likelihood function defined in Equation \eqref{eqn::chi_square_lum_fit}. For the fit, we assumed flat priors for all three parameters: $0.01<\phi_{\star}<5.0$, $-24.0< M_{\star} <-19$, $-2.0< \alpha <-0.01$ (when $\alpha \le -2.0$, the luminosity integration diverges). The sampler was run with 30 walkers, each with 500 steps. We discard the first 150 burn-in steps, which leaves a MCMC chain with a shape of 10500 samples.

Figure \ref{schechter_function_plot} shows the observed excess number per LRG pair in the filament box as a function of apparent magnitude. Error bars in the observations are taken from the diagonal elements in the covariance matrix. The best-fit Schechter function and its uncertainty, calculated by using 50 randomly chosen MCMC samples, are also shown. The observed number at bright magnitudes fluctuates around zero, consistent with the behaviour of luminosity function. Error bars in the CMASS measurements are smaller than the error in LOWZ sample because there are more pairs in the CMASS sample, although the LOWZ sample spans a greater range in magnitude. The best-fit parameters for the LOWZ and CMASS samples are summarised in Table \ref{table_summary_schechter_results}\footnote{Note that in the table, we exclude the Schechter fit results with combined LOWZ and CMASS sample as the $L_{\rm total}$ values for case LOWZ+CMASS are computed by averaging LOWZ and CMASS measurements.}. The nominal values and uncertainties of the three fitted parameters are approximated by the average and standard deviation calculated from MCMC samples. Similarly, we compute the total luminosity and its uncertainties using the correlated parameters from the MCMC samples. The total stellar mass, $M_{\rm stellar, \rm total}$, is computed by multiplying the total luminosity with the observed $M_{\rm stellar, \rm observed}/L_{\rm observed}$ from the 2D map.

Notice that, for the CMASS sample, the degeneracy of the three parameters is significant. This is due to the flux limit of the SDSS imaging and the depth of the CMASS LRGS, by which only the bright part of Schechter function can be observed. This degeneracy could be broken with a deeper dataset, or more simply, using low-redshift sample (LOWZ).

\begin{table*}
\centering
\begin{tabular}{c|cccccccc} 
 \hline
 Sample name & $\overline{z}_{\textrm{pair}}$  & \vtop{\hbox{\strut $M$}\hbox{\strut (1)}}& \vtop{\hbox{\strut ${M_{\textrm{stellar}}}$}\hbox{\strut (2)}}&\vtop{\hbox{\strut ${L_{^{0.1}g}}$}\hbox{\strut (3)}} & \vtop{\hbox{\strut ${L_{^{0.1}r}}$}\hbox{\strut (4)}}  &${M_{\textrm{stellar}}}/M$&$M/L_{^{0.1}r}$& $^{0.1}(g-r)$ \\ [0.5ex] 
 \hline\hline
 $6h^{-1} \textrm{Mpc} \leqslant R_{\textrm{sep}} \leqslant 10 h^{-1} \textrm{Mpc}$ &  &   &  &  &  & & & \\
 \hline
 LOWZ & 0.33 &  4.86$\pm$2.54 & 4.08$\pm$0.77 &  1.70$\pm$0.32 & 1.57$\pm$0.19 &$0.84\pm0.47\%$&310$\pm$166&0.59$\pm$0.24 \\ 
 CMASS & 0.53& 3.63$\pm$1.84 & 2.16$\pm$0.65 & 1.11$\pm$0.33 & 0.84$\pm$0.28  &$0.60\pm0.35\%$&432$\pm$262 & 0.38$\pm$0.45\\
 LOWZ+CMASS & 0.44 & 4.25$\pm$1.57 & 3.12$\pm$0.50 &1.41$\pm$0.23& 1.21$\pm$0.15 &$0.73\pm0.30\%$ &351$\pm$137&0.51$\pm$0.22\\
 \hline
 $3h^{-1} \textrm{Mpc} \leqslant R_{\textrm{sep}} \leqslant 5 h^{-1} \textrm{Mpc}$ &  &   &  &  &  & & & \\
 \hline
 LOWZ & 0.30 &  4.35$\pm$1.35 & 1.78$\pm$0.74&  0.56$\pm$0.23 & 0.73$\pm$0.26 &$0.41\pm0.21\%$&596$\pm$282&0.97$\pm$0.59\\
 CMASS & 0.50 &  2.60$\pm$1.18 & 1.69$\pm$0.43&  0.65$\pm$0.12 & 0.62$\pm$0.11 &$0.65\pm0.34\%$&419$\pm$204&0.63$\pm$0.28\\
 LOWZ+CMASS & 0.40 &  3.48$\pm$0.90 & 1.74$\pm$0.43&  0.61$\pm$0.13 & 0.68$\pm$0.14 &$0.50\pm0.18\%$&512$\pm$169&0.80$\pm$0.32
 \\ 
 \hline
\end{tabular}
\caption{Summary of the measured filament properties. The results of $M$,$M_{\rm stellar}$,$L_{^{0.1}g}$ and $L_{^{0.1}r}$ for the LOWZ + CMASS sample are computed from the simple average between LOWZ and CMASS sample. Note: (1): total filament mass in units of $10^{13}M_{\odot}$ determined from weak lensing (Section \ref{ssec::lensing_results}). (2): total filament stellar mass in units of $10^{11}M_{\odot}$ calculated using the observed stellar-mass-to-light ratio and the total light. The (3),(4): total luminosity in units of $10^{11}L_{\odot}$ for the $^{0.1}g$- and $^{0.1}r$-bands, determined by integrating the Schechter function. The best-fit parameters of the Schechter function are listed in Table \ref{table_summary_schechter_results}. For comparison, the last three rows show the results for the LRG pair separation criteria of \citet{new_filament_paper} .}

\label{table_summary_results}
\end{table*}

\begin{table*}
  \begin{tabular}{c|ccc|ccc}
    \hline
    \multirow{2}{*}{\vtop{\hbox{\strut Sample name}\hbox{\strut }}}
    &
      \multicolumn{3}{c}{$^{0.1}r$ band fit} &
      \multicolumn{3}{c}{$^{0.1}g$ band fit}\\
    \hline
    \hline
    & \vtop{\hbox{\strut $n_{\star}$}} & $\alpha$&$M_{\star, ^{0.1}r}$& \vtop{\hbox{\strut $n_{\star}$}} & $\alpha$& $M_{\star, ^{0.1}g}$ \\
    \hline
    $6h^{-1} \textrm{Mpc} \leqslant R_{\textrm{sep}} \leqslant 10 h^{-1} \textrm{Mpc}$ &  &   &  &  &  & \\
    \hline
    LOWZ & 4.24$\pm$0.77 & -1.05$\pm$0.09 &-21.67$\pm$0.14 & 4.72$\pm$1.48& -1.38$\pm$0.17 &-20.96$\pm$0.26 \\
    CMASS & 2.33$\pm$1.15 & -0.83$\pm$0.41 & -21.63$\pm$0.38 & 2.78$\pm$1.27 & -1.20$\pm$0.36 & -21.08$\pm$0.39 \\
    \hline
    $3h^{-1} \textrm{Mpc} \leqslant R_{\textrm{sep}} \leqslant 5 h^{-1} \textrm{Mpc}$ &  &   &  &  &  & \\
    \hline
    LOWZ & 3.17$\pm$1.40 & -0.72$\pm$0.27 &-21.17$\pm$0.29 & 3.73$\pm$1.83& -0.81$\pm$0.52 &-20.00$\pm$0.32 \\
    CMASS & 2.30$\pm$0.48 & -1.00 (fixed) & -21.34$\pm$0.12 & 4.06$\pm$0.79 & -1.00 (fixed) & -20.09$\pm$0.10 \\
    \hline
  \end{tabular}
  \caption{Best-fit parameters of Schechter function for different catalogues in $^{0.1}r$ and $^{0.1}g$ band. Best fits are obtained from MCMC and the uncertainties are calculated from MCMC chains.}
 \label{table_summary_schechter_results}
 \end{table*}

Finally, we consider the stellar populations of filament galaxies, as estimated via their mean colour, as determined from the luminosities in different bands, as follows:
\begin{equation}\label{eqn::colour_difference}
^{0.1}(g-r) = -2.5\log_{10}\bigg(\frac{L_{^{0.1}g,\textrm{total}}}{L_{^{0.1}r,\textrm{total}}}\bigg) + M_{\odot, ^{0.1}g} - M_{\odot, ^{0.1}r},
\end{equation}
where the luminosity is measured in solar units, the magnitude of the Sun are  $M_{\odot, ^{0.1}g}$ = 5.45 and $M_{\odot, ^{0.1}r}$ = 4.76 \citep{color_difference_Blanton}, and  uncertainties in the colour are propagated from the total luminosity errors. The results for different sample catalogues are summarised in the last column in Table \ref{table_summary_results}. 

\begin{figure*}
    \begin{minipage}[b]{0.5\textwidth}
        \centering
        \includegraphics[width=1.1\linewidth]{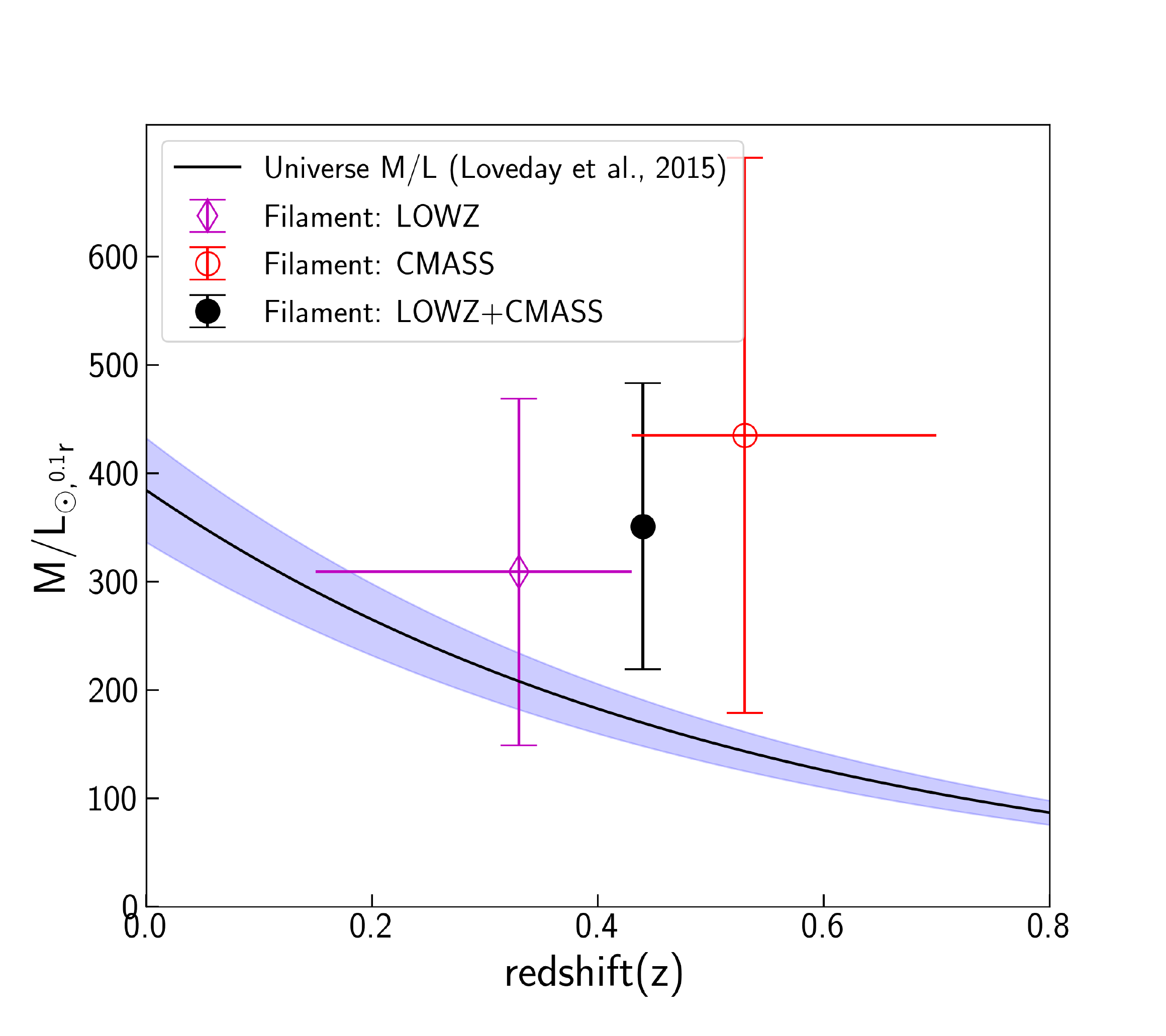}
    \end{minipage}\begin{minipage}[b]{0.5\textwidth}
        \centering
        \includegraphics[width=1.1\linewidth]{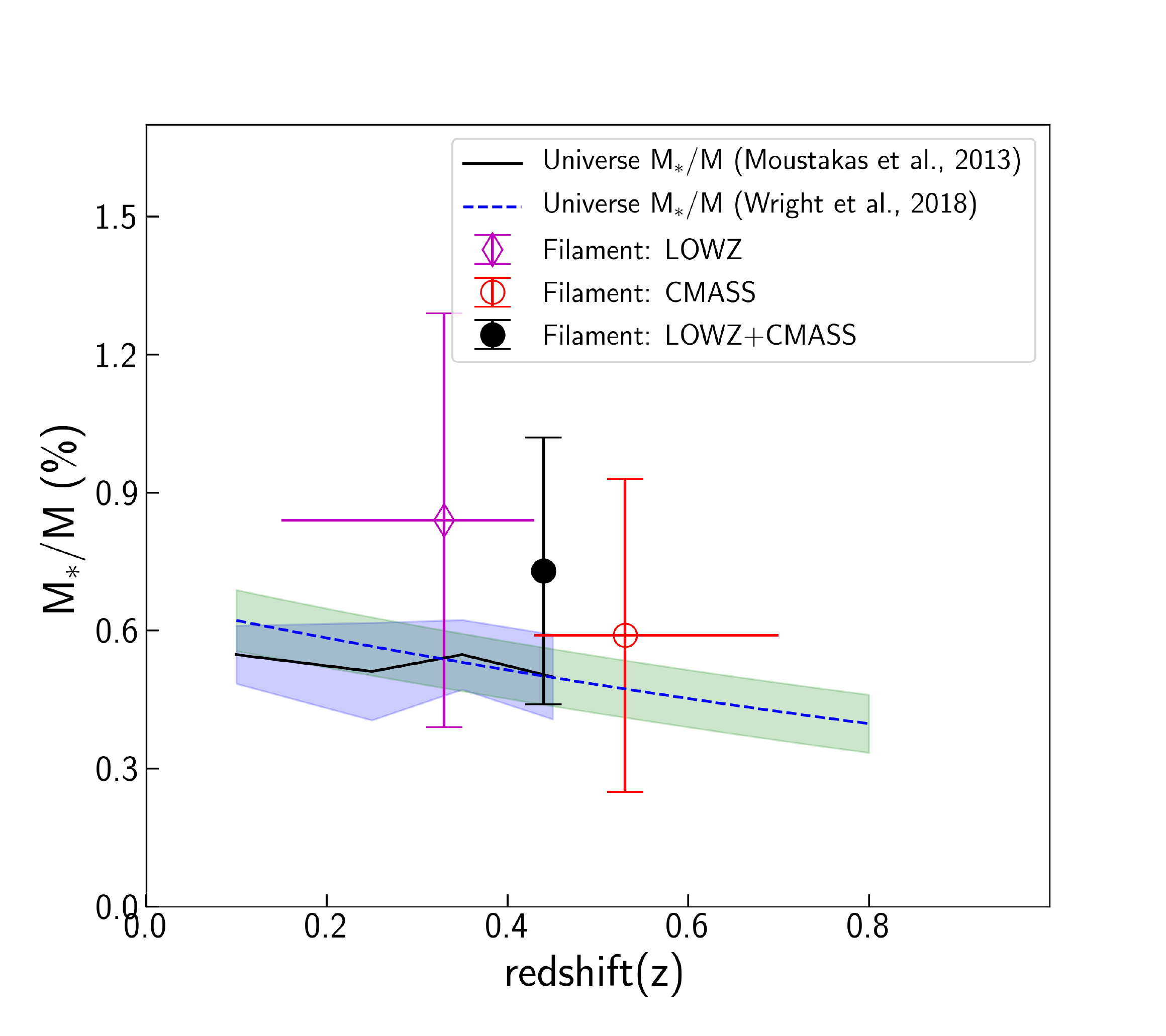}
    \end{minipage}%
\caption{Left panel: The filament $M/L$ as a function of redshift. The black solid line with 1-$\sigma$ shaded region shows the universe M/L value as a function of redshift predicted by \citet{M_L_evolution}. The thin magenta diamond and hollow red circle with error bars are the filament measurements from two independent samples (LOWZ and CMASS), where the horizontal error bars indicate the redshift coverage. The black circle with black errorbar shows the results from averaging LOWZ and CMASS. Right panel: The filament $M_{\rm stellar}$/M as a function of redshift. The black solid line with shaded region shows the universal average $M_{\rm stellar}/M$ value as a function of redshift from \citet{stellar_mass_ratio_paper}, and the blue dashed line shows the same quantity derived from \citet{wright_M_stellar_M_comparison}.}
\label{ratio_results}
\end{figure*}

\begin{figure}
    \centering
        \includegraphics[width=1.0\linewidth]{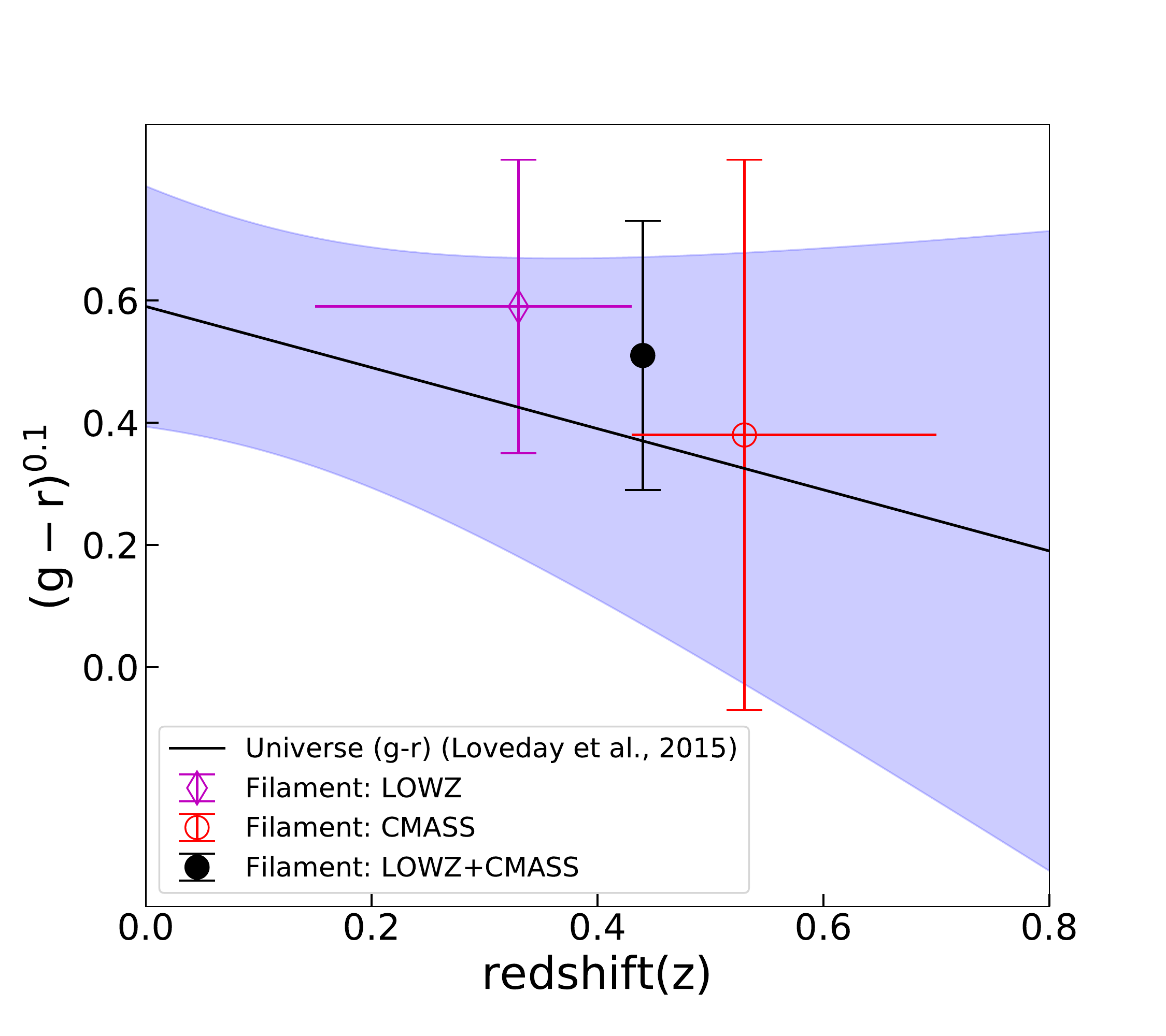}
\caption{Filament color as a function of redshift. The black solid line with 1-$\sigma$ shaded region shows the universe (g-r) value as a function of redshift predicted by \citet{Loverday_color_difference}. Magenta thin diamond and red hollow circle with errorbars are the colour values of filaments measured from LOWZ and CMASS. Black point with black errorbar shows the results from the combined samples.}
\label{g_r_results}
\end{figure}

\section{Discussion}\label{sec::discussion}

With the measurements of $M_{\rm total}$, $M_{\rm stellar, \rm total}$ and $L_{\rm total}$ obtained in a consistent way, we calculate the total-mass-to-light ratio and the stellar-mass-to-total-mass ratio. The results are in Table \ref{table_summary_results}, and shown in Figure \ref{ratio_results}. Within the uncertainties there is no significant evolution of either ratio from $z=0.33$ to $z=0.53$.

First, we compare our results with the universal average mass-to-light ratio from \citet{M_L_evolution}. Using GAMA-II galaxy sample, they estimated the comoving luminosity density evolution in the $^{0.1}r$ band over the redshift range $0.002<z<0.65$. Based on their results, the calculated $M/L$ evolution as a function of redshift is shown in the left panel of Figure \ref{ratio_results}, where we have adopted a comoving matter density, for $\Omega_{m,0}=0.3$, of $\rho_{\rm m} = 4.09\times10^{10} M_{\odot} \rm Mpc^{-3}$. At $z  = 0.33$, their model predicts $M/L = 208\pm26$, and at $z = 0.53$ the ratio is 143$\pm$18. Our result is slightly higher than the universal average suggesting filaments are darker than the Universe as a whole. 

The $M_{\rm stellar}/M$ as a function of redshift is illustrated in the right panel of Figure \ref{ratio_results}. For comparison, we also over-plot the results obtained by \citet{stellar_mass_ratio_paper} and \citet{wright_M_stellar_M_comparison}. As can be seen from this figure, the stellar mass fraction values and their redshift evolution obtained in our study are consistent with the results found in both of these papers. 

With regards to the colour of the stellar populations of filament galaxies, due to the large uncertainties, we find no statistically significant evolution over the redshift range considered. We can compare our results with the cosmic-average colour values obtained from \citet{color_difference_Blanton}, where they measure the $^{0.1}(g-r)$ value by fitting the luminosity function at $z$ = 0.1 using SDSS galaxies. In their study, the $^{0.1}(g-r)$ value obtained at $z$ = 0.1 is $0.73\pm0.04$. In order estimate the evolution of the colour as a function of redshift, we use the LF evolution parameters of  \citet{Loverday_color_difference}. The colour, $^{0.1}\widetilde{(g-r)}$\footnote{To obtain an estimate on how the colour changes with redshift, we ignore all uncertainties in the parameter fits and only use the nominal values.}, of the universe changes from 0.43 at $z$ = 0.33, 0.37 at $z$ = 0.44, to 0.33 at $z$ = 0.53. Using the best-fit parametric luminosity function from \citet{Loverday_color_difference}, the colour as a function of redshift is shown in Figure \ref{g_r_results}, where we also over-plot the colour measurements from filaments obtained by this study. The colours of filament galaxies are consistent with the universal average colour, given the uncertainties.

In summary, the picture that emerges is that, within the uncertainties, the mass-to-light ratios, stellar-to-total mass fraction and the colours are statistically consistent with the cosmic averages. One interpretation of this is that both the stellar mass fractions as a function of halo mass, and the halo abundances in the filament environment are close to the cosmic averages. Due to the large error bars, it is hard to put strong constraints on the redshift dependence of these quantities. 

\section{Conclusion}\label{sec::conclusion}

The key results of this paper are the first measurements of the mass-to-light ratio and stellar mass fraction of filaments in the cosmic web. To achieve this goal, following the procedure outlined in \citet{Seth_filament_paper}, we reproduced the masses of stacked filaments between LRG pairs. Using a consistent scheme for stacking, we then produced the 2D projected luminosity and stellar mass density maps of the stacked filaments. We show that the luminosity maps of the filaments have a characteristic cylindrical radius $r_c = (0.47\pm0.07) h^{-1}\rm Mpc$. We conducted these analyses for two independent LRG samples (LOWZ and CMASS) at different redshifts, but over the limited redshift range, there is no significant evolution of the $M_{\rm total}/L_{\rm total}$ or $M_{\rm stellar}/M_{\rm total}$.  The $M_{\rm stellar}/M_{\rm total}$ values and average colours of galaxies in filaments are consistent with the universal averages. In summary, the filaments of the cosmic web, although dominated by the dark matter, are not entirely dark.

The uncertainties remain large for this analysis, given current data. The advent of other surveys could do help to provide more accurate and precise measurements. For instance, the recently-completed Dark Energy Survey \citep[DES:][]{DES} and ongoing surveys such as the Hyper Suprime-Cam Subaru Strategic Program \citep[HSC-SSP:][]{HSC} and the Canada-France Imaging Survey \citep[CFIS:][]{CFIS} will provide deep, wide-field shape catalogues. Surveys such as SDSS-IV Extended Baryon Oscillation Spectroscopic Survey \citep[eBOSS:][]{eboss_paper} and DESI survey \citep[DESI:][]{DESI_survey} will measure more galaxy redshifts, mapping large-scale structure in deeper and larger volumes. Space missions such as Euclid \citep{euclid_paper} and WFIRST \citep{wfirst_paper} will improve the weak lensing measurements by orders of magnitude while proposed spectroscopic missions such as ATLAS probe \citep{three_point_correlation_function_prediction} could map the cosmic web at high resolution. With the increase in the accuracy and completeness of different surveys, we hope to provide more reliable and complete measurements of the cosmic web.   

\section*{Acknowledgements}

MH and NA acknowledge an NSERC Discovery grant. This work was supported by the University of Waterloo, Natural Sciences and Engineering Research Council of Canada (NSERC), and the Perimeter Institute for Theoretical Physics. Research at the Perimeter Institute is supported by the Government of Canada through Industry Canada, and by the Province of Ontario through the Ministry of Research and Innovation.

This work is based on observations obtained with MegaPrime/MegaCam, a joint project of CFHT and CEA/IRFU, at the Canada-FranceHawaii Telescope (CFHT) which is operated by the National Research Council (NRC) of Canada, the Institut National des Science de l'Univers of the Centre National de la Recherche Scientifique (CNRS) of France, and the University of Hawaii. This work is based in part on data products produced at Terapix available at the Canadian Astronomy Data Centre as part of the Canada-FranceHawaii Telescope Legacy Survey, a collaborative project of NRC and CNRS.

Funding for SDSS-III has been provided by the Alfred P.
Sloan Foundation, the Participating Institutions, the National Science
Foundation, and the U.S. Department of Energy Office of Science.
The SDSS-III website is \url{http://www.sdss3.org/}.

SDSS-III is managed by the Astrophysical Research Consortium for the Participating Institutions of the SDSS-III Collaboration including the University of Arizona, the Brazilian Participation Group, Brookhaven National Laboratory, University of Cambridge, Carnegie Mellon University, University of Florida, the French Participation Group, the German Participation Group, Harvard University, the Instituto de Astrofisica de Canarias, the Michigan State/Notre Dame/JINA Participation Group, Johns Hopkins University, Lawrence Berkeley National Laboratory, Max Planck Institute for Astrophysics, Max Planck Institute for Extraterrestrial Physics, New Mexico State University, New York University, Ohio State University, Pennsylvania State University, University of Portsmouth, Princeton University, the Spanish Participation Group, University of Tokyo, University of Utah, Vanderbilt University, University of Virginia, University of Washington, and Yale University.

\section*{Data availability}
The data underlying this article are available in Zenodo, at \url{https://doi.org/10.5281/zenodo.3951947}.

\bibliographystyle{mnras}
\bibliography{draft}

\begin{thebibliography}{}
\makeatletter
\relax
\def\mn@urlcharsother{\let\do\@makeother \do\$\do\&\do\#\do\^\do\_\do\%\do\~}
\def\mn@doi{\begingroup\mn@urlcharsother \@ifnextchar [ {\mn@doi@}
  {\mn@doi@[]}}
\def\mn@doi@[#1]#2{\def\@tempa{#1}\ifx\@tempa\@empty \href
  {http://dx.doi.org/#2} {doi:#2}\else \href {http://dx.doi.org/#2} {#1}\fi
  \endgroup}
\def\mn@eprint#1#2{\mn@eprint@#1:#2::\@nil}
\def\mn@eprint@arXiv#1{\href {http://arxiv.org/abs/#1} {{\tt arXiv:#1}}}
\def\mn@eprint@dblp#1{\href {http://dblp.uni-trier.de/rec/bibtex/#1.xml}
  {dblp:#1}}
\def\mn@eprint@#1:#2:#3:#4\@nil{\def\@tempa {#1}\def\@tempb {#2}\def\@tempc
  {#3}\ifx \@tempc \@empty \let \@tempc \@tempb \let \@tempb \@tempa \fi \ifx
  \@tempb \@empty \def\@tempb {arXiv}\fi \@ifundefined
  {mn@eprint@\@tempb}{\@tempb:\@tempc}{\expandafter \expandafter \csname
  mn@eprint@\@tempb\endcsname \expandafter{\@tempc}}}

\bibitem[\protect\citeauthoryear{{Aihara} et~al.,}{{Aihara} et~al.}{2018}]{HSC}
{Aihara} H.,  et~al., 2018, \mn@doi [\pasj] {10.1093/pasj/psx066}, \href
  {https://ui.adsabs.harvard.edu/abs/2018PASJ...70S...4A} {70, S4}

\bibitem[\protect\citeauthoryear{{Beck}, {Dobos}, {Budav{\'a}ri}, {Szalay}  \&
  {Csabai}}{{Beck} et~al.}{2016}]{photoz_SDSS}
{Beck} R.,  {Dobos} L.,  {Budav{\'a}ri} T.,  {Szalay} A.~S.,   {Csabai} I.,
  2016, \mn@doi [\mnras] {10.1093/mnras/stw1009}, \href
  {https://ui.adsabs.harvard.edu/abs/2016MNRAS.460.1371B} {460, 1371}

\bibitem[\protect\citeauthoryear{{Behroozi}, {Conroy}  \&
  {Wechsler}}{{Behroozi} et~al.}{2010}]{Behroozi_abundance_matching_paper_1}
{Behroozi} P.~S.,  {Conroy} C.,   {Wechsler} R.~H.,  2010, \mn@doi [\apj]
  {10.1088/0004-637X/717/1/379}, \href
  {https://ui.adsabs.harvard.edu/abs/2010ApJ...717..379B} {717, 379}

\bibitem[\protect\citeauthoryear{{Behroozi}, {Wechsler}  \&
  {Conroy}}{{Behroozi} et~al.}{2013}]{Behroozi_abundance_matching_paper_2}
{Behroozi} P.~S.,  {Wechsler} R.~H.,   {Conroy} C.,  2013, \mn@doi [\apj]
  {10.1088/0004-637X/770/1/57}, \href
  {https://ui.adsabs.harvard.edu/abs/2013ApJ...770...57B} {770, 57}

\bibitem[\protect\citeauthoryear{{Blanton} et~al.,}{{Blanton}
  et~al.}{2003}]{color_difference_Blanton}
{Blanton} M.~R.,  et~al., 2003, \mn@doi [\apj] {10.1086/375776}, \href
  {https://ui.adsabs.harvard.edu/abs/2003ApJ...592..819B} {592, 819}

\bibitem[\protect\citeauthoryear{{Bond}, {Kofman}  \& {Pogosyan}}{{Bond}
  et~al.}{1996}]{Bond_web}
{Bond} J.~R.,  {Kofman} L.,   {Pogosyan} D.,  1996, \mn@doi [Nature]
  {10.1038/380603a0}, \href
  {https://ui.adsabs.harvard.edu/abs/1996Natur.380..603B} {380, 603}

\bibitem[\protect\citeauthoryear{{Clampitt}, {Jain}, {Takada}  \&
  {Miyatake}}{{Clampitt} et~al.}{2014}]{Clampitt_2014}
{Clampitt} J.,  {Jain} B.,  {Takada} M.,   {Miyatake} H.,  2014, arXiv
  e-prints, p. arXiv:1402.3302v1

\bibitem[\protect\citeauthoryear{{Clampitt}, {Miyatake}, {Jain}  \&
  {Takada}}{{Clampitt} et~al.}{2016}]{clampitt_fil_detection}
{Clampitt} J.,  {Miyatake} H.,  {Jain} B.,   {Takada} M.,  2016, \mn@doi
  [\mnras] {10.1093/mnras/stw142}, \href
  {https://ui.adsabs.harvard.edu/abs/2016MNRAS.457.2391C} {457, 2391}

\bibitem[\protect\citeauthoryear{{Colberg}, {Krughoff}  \&
  {Connolly}}{{Colberg} et~al.}{2005}]{Intercluster_filaments_in_Universe}
{Colberg} J.~M.,  {Krughoff} K.~S.,   {Connolly} A.~J.,  2005, \mn@doi [\mnras]
  {10.1111/j.1365-2966.2005.08897.x}, \href
  {https://ui.adsabs.harvard.edu/abs/2005MNRAS.359..272C} {359, 272}

\bibitem[\protect\citeauthoryear{{Colless} et~al.,}{{Colless}
  et~al.}{2003}]{2dFGRS_fil}
{Colless} M.,  et~al., 2003, arXiv e-prints, \href
  {https://ui.adsabs.harvard.edu/abs/2003astro.ph..6581C} {pp
  astro--ph/0306581}

\bibitem[\protect\citeauthoryear{{DESI Collaboration} et~al.,}{{DESI
  Collaboration} et~al.}{2016}]{DESI_survey}
{DESI Collaboration} et~al., 2016, arXiv e-prints, \href
  {https://ui.adsabs.harvard.edu/abs/2016arXiv161100036D} {p. arXiv:1611.00036}

\bibitem[\protect\citeauthoryear{{Dawson} et~al.,}{{Dawson}
  et~al.}{2013}]{Dawson_paper}
{Dawson} K.~S.,  et~al., 2013, \mn@doi [\aj] {10.1088/0004-6256/145/1/10},
  \href {https://ui.adsabs.harvard.edu/abs/2013AJ....145...10D} {145, 10}

\bibitem[\protect\citeauthoryear{{Dawson} et~al.,}{{Dawson}
  et~al.}{2016}]{eboss_paper}
{Dawson} K.~S.,  et~al., 2016, \mn@doi [\aj] {10.3847/0004-6256/151/2/44},
  \href {https://ui.adsabs.harvard.edu/abs/2016AJ....151...44D} {151, 44}

\bibitem[\protect\citeauthoryear{{Dietrich}, {Werner}, {Clowe}, {Finoguenov},
  {Kitching}, {Miller}  \& {Simionescu}}{{Dietrich}
  et~al.}{2012}]{Abell_cluster_lensing_detection}
{Dietrich} J.~P.,  {Werner} N.,  {Clowe} D.,  {Finoguenov} A.,  {Kitching} T.,
  {Miller} L.,   {Simionescu} A.,  2012, \mn@doi [\nat] {10.1038/nature11224},
  \href {https://ui.adsabs.harvard.edu/abs/2012Natur.487..202D} {487, 202}

\bibitem[\protect\citeauthoryear{{Dobos}, {Csabai}, {Yip}, {Budav{\'a}ri},
  {Wild}  \& {Szalay}}{{Dobos} et~al.}{2012}]{An_atlas_of_composite_spectra}
{Dobos} L.,  {Csabai} I.,  {Yip} C.-W.,  {Budav{\'a}ri} T.,  {Wild} V.,
  {Szalay} A. e.~S.,  2012, \mn@doi [\mnras]
  {10.1111/j.1365-2966.2011.20109.x}, \href
  {https://ui.adsabs.harvard.edu/abs/2012MNRAS.420.1217D} {420, 1217}

\bibitem[\protect\citeauthoryear{{Eisenstein} et~al.,}{{Eisenstein}
  et~al.}{2001}]{LRG_selected}
{Eisenstein} D.~J.,  et~al., 2001, \mn@doi [\aj] {10.1086/323717}, \href
  {https://ui.adsabs.harvard.edu/abs/2001AJ....122.2267E} {122, 2267}

\bibitem[\protect\citeauthoryear{{Eisenstein} et~al.,}{{Eisenstein}
  et~al.}{2011}]{Eisenstein_boss_paper}
{Eisenstein} D.~J.,  et~al., 2011, \mn@doi [\aj] {10.1088/0004-6256/142/3/72},
  \href {https://ui.adsabs.harvard.edu/abs/2011AJ....142...72E} {142, 72}

\bibitem[\protect\citeauthoryear{{Epps} \& {Hudson}}{{Epps} \&
  {Hudson}}{2017}]{Seth_filament_paper}
{Epps} S.~D.,  {Hudson} M.~J.,  2017, \mn@doi [\mnras] {10.1093/mnras/stx517},
  \href {https://ui.adsabs.harvard.edu/abs/2017MNRAS.468.2605E} {468, 2605}

\bibitem[\protect\citeauthoryear{{Erben} et~al.,}{{Erben}
  et~al.}{2013}]{CFHTLens_general_Erben}
{Erben} T.,  et~al., 2013, \mn@doi [\mnras] {10.1093/mnras/stt928}, \href
  {https://ui.adsabs.harvard.edu/abs/2013MNRAS.433.2545E} {433, 2545}

\bibitem[\protect\citeauthoryear{{Foreman-Mackey}, {Hogg}, {Lang}  \&
  {Goodman}}{{Foreman-Mackey} et~al.}{2013}]{emcee_paper}
{Foreman-Mackey} D.,  {Hogg} D.~W.,  {Lang} D.,   {Goodman} J.,  2013, \mn@doi
  [\pasp] {10.1086/670067}, \href
  {https://ui.adsabs.harvard.edu/abs/2013PASP..125..306F} {125, 306}

\bibitem[\protect\citeauthoryear{{Gil-Mar{\'\i}n}, {Percival}, {Verde},
  {Brownstein}, {Chuang}, {Kitaura}, {Rodr{\'\i}guez-Torres}  \&
  {Olmstead}}{{Gil-Mar{\'\i}n}
  et~al.}{2017}]{LRG_halo_bias_LOWZ_CMASS_separate}
{Gil-Mar{\'\i}n} H.,  {Percival} W.~J.,  {Verde} L.,  {Brownstein} J.~R.,
  {Chuang} C.-H.,  {Kitaura} F.-S.,  {Rodr{\'\i}guez-Torres} S.~A.,
  {Olmstead} M.~D.,  2017, \mn@doi [\mnras] {10.1093/mnras/stw2679}, \href
  {https://ui.adsabs.harvard.edu/abs/2017MNRAS.465.1757G} {465, 1757}

\bibitem[\protect\citeauthoryear{{Guzzo} et~al.,}{{Guzzo}
  et~al.}{2014}]{VIPER_fil}
{Guzzo} L.,  et~al., 2014, \mn@doi [\aap] {10.1051/0004-6361/201321489}, \href
  {https://ui.adsabs.harvard.edu/abs/2014A&A...566A.108G} {566, A108}

\bibitem[\protect\citeauthoryear{{He}, {Alam}, {Ferraro}, {Chen}  \& {Ho}}{{He}
  et~al.}{2018}]{cmb_filament}
{He} S.,  {Alam} S.,  {Ferraro} S.,  {Chen} Y.-C.,   {Ho} S.,  2018, \mn@doi
  [Nature Astronomy] {10.1038/s41550-018-0426-z}, \href
  {https://ui.adsabs.harvard.edu/abs/2018NatAs...2..401H} {2, 401}

\bibitem[\protect\citeauthoryear{{Heymans} et~al.,}{{Heymans}
  et~al.}{2012}]{CFHTLens_general_Heymans}
{Heymans} C.,  et~al., 2012, \mn@doi [\mnras]
  {10.1111/j.1365-2966.2012.21952.x}, \href
  {https://ui.adsabs.harvard.edu/abs/2012MNRAS.427..146H} {427, 146}

\bibitem[\protect\citeauthoryear{{Higuchi}, {Oguri}  \& {Shirasaki}}{{Higuchi}
  et~al.}{2014}]{Statistical_properties_of_filaments_in_weak_gravitational_lensing}
{Higuchi} Y.,  {Oguri} M.,   {Shirasaki} M.,  2014, \mn@doi [\mnras]
  {10.1093/mnras/stu583}, \href
  {https://ui.adsabs.harvard.edu/abs/2014MNRAS.441..745H} {441, 745}

\bibitem[\protect\citeauthoryear{{Hildebrandt} et~al.,}{{Hildebrandt}
  et~al.}{2012}]{CFHTLens_redshift}
{Hildebrandt} H.,  et~al., 2012, \mn@doi [\mnras]
  {10.1111/j.1365-2966.2012.20468.x}, \href
  {https://ui.adsabs.harvard.edu/abs/2012MNRAS.421.2355H} {421, 2355}

\bibitem[\protect\citeauthoryear{{Huchra} et~al.,}{{Huchra}
  et~al.}{2012}]{2MASS_fil}
{Huchra} J.~P.,  et~al., 2012, \mn@doi [\apjs] {10.1088/0067-0049/199/2/26},
  \href {https://ui.adsabs.harvard.edu/abs/2012ApJS..199...26H} {199, 26}

\bibitem[\protect\citeauthoryear{{Hudson} et~al.,}{{Hudson}
  et~al.}{2015}]{lensing_source_galaxy_selection}
{Hudson} M.~J.,  et~al., 2015, \mn@doi [\mnras] {10.1093/mnras/stu2367}, \href
  {https://ui.adsabs.harvard.edu/abs/2015MNRAS.447..298H} {447, 298}

\bibitem[\protect\citeauthoryear{{Ibata} et~al.,}{{Ibata} et~al.}{2017}]{CFIS}
{Ibata} R.~A.,  et~al., 2017, \mn@doi [\apj] {10.3847/1538-4357/aa855c}, \href
  {https://ui.adsabs.harvard.edu/abs/2017ApJ...848..128I} {848, 128}

\bibitem[\protect\citeauthoryear{{Jauzac} et~al.,}{{Jauzac}
  et~al.}{2012}]{A_weak_lensing_mass_reconstruction_filament_MACS}
{Jauzac} M.,  et~al., 2012, \mn@doi [\mnras]
  {10.1111/j.1365-2966.2012.21966.x}, \href
  {https://ui.adsabs.harvard.edu/abs/2012MNRAS.426.3369J} {426, 3369}

\bibitem[\protect\citeauthoryear{{Kaiser} \& {Squires}}{{Kaiser} \&
  {Squires}}{1993}]{Kaiser_Squires_lensing}
{Kaiser} N.,  {Squires} G.,  1993, \mn@doi [\apj] {10.1086/172297}, \href
  {https://ui.adsabs.harvard.edu/abs/1993ApJ...404..441K} {404, 441}

\bibitem[\protect\citeauthoryear{{Kazin} et~al.,}{{Kazin}
  et~al.}{2010}]{SDSS_DR7_LRG}
{Kazin} E.~A.,  et~al., 2010, \mn@doi [\apj] {10.1088/0004-637X/710/2/1444},
  \href {https://ui.adsabs.harvard.edu/abs/2010ApJ...710.1444K} {710, 1444}

\bibitem[\protect\citeauthoryear{{Kondo}, {Miyatake}, {Shirasaki}, {Sugiyama}
  \& {Nishizawa}}{{Kondo} et~al.}{2019}]{Kondo_stacking_fil}
{Kondo} H.,  {Miyatake} H.,  {Shirasaki} M.,  {Sugiyama} N.,   {Nishizawa}
  A.~J.,  2019, arXiv e-prints, \href
  {https://ui.adsabs.harvard.edu/abs/2019arXiv190508991K} {p. arXiv:1905.08991}

\bibitem[\protect\citeauthoryear{{Kravtsov}, {Vikhlinin}  \&
  {Meshcheryakov}}{{Kravtsov} et~al.}{2018}]{Kravtsov_2018_SF_efficiency}
{Kravtsov} A.~V.,  {Vikhlinin} A.~A.,   {Meshcheryakov} A.~V.,  2018, \mn@doi
  [Astronomy Letters] {10.1134/S1063773717120015}, \href
  {https://ui.adsabs.harvard.edu/abs/2018AstL...44....8K} {44, 8}

\bibitem[\protect\citeauthoryear{{Laureijs} et~al.,}{{Laureijs}
  et~al.}{2011}]{euclid_paper}
{Laureijs} R.,  et~al., 2011, arXiv e-prints, \href
  {https://ui.adsabs.harvard.edu/abs/2011arXiv1110.3193L} {p. arXiv:1110.3193}

\bibitem[\protect\citeauthoryear{{Libeskind}, {Guo}, {Tempel}  \&
  {Ibata}}{{Libeskind} et~al.}{2016}]{lopsided_distribution_satellite}
{Libeskind} N.~I.,  {Guo} Q.,  {Tempel} E.,   {Ibata} R.,  2016, \mn@doi [\apj]
  {10.3847/0004-637X/830/2/121}, \href
  {https://ui.adsabs.harvard.edu/abs/2016ApJ...830..121L} {830, 121}

\bibitem[\protect\citeauthoryear{{Libeskind} et~al.,}{{Libeskind}
  et~al.}{2018}]{Tracing_the_cosmic_web}
{Libeskind} N.~I.,  et~al., 2018, \mn@doi [\mnras] {10.1093/mnras/stx1976},
  \href {https://ui.adsabs.harvard.edu/abs/2018MNRAS.473.1195L} {473, 1195}

\bibitem[\protect\citeauthoryear{{Loveday} et~al.,}{{Loveday}
  et~al.}{2012}]{Loverday_color_difference}
{Loveday} J.,  et~al., 2012, \mn@doi [\mnras]
  {10.1111/j.1365-2966.2011.20111.x}, \href
  {https://ui.adsabs.harvard.edu/abs/2012MNRAS.420.1239L} {420, 1239}

\bibitem[\protect\citeauthoryear{{Loveday} et~al.,}{{Loveday}
  et~al.}{2015}]{M_L_evolution}
{Loveday} J.,  et~al., 2015, \mn@doi [\mnras] {10.1093/mnras/stv1013}, \href
  {https://ui.adsabs.harvard.edu/abs/2015MNRAS.451.1540L} {451, 1540}

\bibitem[\protect\citeauthoryear{{Mandelbaum}, {Seljak}, {Cool}, {Blanton},
  {Hirata}  \& {Brinkmann}}{{Mandelbaum}
  et~al.}{2006}]{LRG_good_aporxy_rich_group}
{Mandelbaum} R.,  {Seljak} U.,  {Cool} R.~J.,  {Blanton} M.,  {Hirata} C.~M.,
  {Brinkmann} J.,  2006, \mn@doi [\mnras] {10.1111/j.1365-2966.2006.10906.x},
  \href {https://ui.adsabs.harvard.edu/abs/2006MNRAS.372..758M} {372, 758}

\bibitem[\protect\citeauthoryear{{Marinoni} \& {Hudson}}{{Marinoni} \&
  {Hudson}}{2002}]{m_L_relation_Marinoni_hudson}
{Marinoni} C.,  {Hudson} M.~J.,  2002, \mn@doi [\apj] {10.1086/339319}, \href
  {https://ui.adsabs.harvard.edu/abs/2002ApJ...569..101M} {569, 101}

\bibitem[\protect\citeauthoryear{{Mead}, {King}  \& {McCarthy}}{{Mead}
  et~al.}{2010}]{Mead_filament_paper}
{Mead} J. M.~G.,  {King} L.~J.,   {McCarthy} I.~G.,  2010, \mn@doi [\mnras]
  {10.1111/j.1365-2966.2009.15840.x}, \href
  {https://ui.adsabs.harvard.edu/abs/2010MNRAS.401.2257M} {401, 2257}

\bibitem[\protect\citeauthoryear{{Miller} et~al.,}{{Miller}
  et~al.}{2013}]{ellipticity_CFHTLens}
{Miller} L.,  et~al., 2013, \mn@doi [\mnras] {10.1093/mnras/sts454}, \href
  {https://ui.adsabs.harvard.edu/abs/2013MNRAS.429.2858M} {429, 2858}

\bibitem[\protect\citeauthoryear{{Miyatake} et~al.,}{{Miyatake}
  et~al.}{2015}]{CFHT_lens_region_cut}
{Miyatake} H.,  et~al., 2015, \mn@doi [\apj] {10.1088/0004-637X/806/1/1}, \href
  {https://ui.adsabs.harvard.edu/abs/2015ApJ...806....1M} {806, 1}

\bibitem[\protect\citeauthoryear{{More}, {van den Bosch}, {Cacciato}, {Mo},
  {Yang}  \& {Li}}{{More} et~al.}{2009}]{More_satellite_dynamics_SDSS}
{More} S.,  {van den Bosch} F.~C.,  {Cacciato} M.,  {Mo} H.~J.,  {Yang} X.,
  {Li} R.,  2009, \mn@doi [\mnras] {10.1111/j.1365-2966.2008.14095.x}, \href
  {https://ui.adsabs.harvard.edu/abs/2009MNRAS.392..801M} {392, 801}

\bibitem[\protect\citeauthoryear{{Moustakas} et~al.,}{{Moustakas}
  et~al.}{2013}]{stellar_mass_ratio_paper}
{Moustakas} J.,  et~al., 2013, \mn@doi [\apj] {10.1088/0004-637X/767/1/50},
  \href {https://ui.adsabs.harvard.edu/abs/2013ApJ...767...50M} {767, 50}

\bibitem[\protect\citeauthoryear{{Parejko} et~al.,}{{Parejko}
  et~al.}{2013}]{LRG_halo_center}
{Parejko} J.~K.,  et~al., 2013, \mn@doi [Monthly Notices of the Royal
  Astronomical Society] {10.1093/mnras/sts314}, \href
  {https://ui.adsabs.harvard.edu/abs/2013MNRAS.429...98P} {429, 98}

\bibitem[\protect\citeauthoryear{{Parker}, {Hudson}, {Carlberg}  \&
  {Hoekstra}}{{Parker} et~al.}{2005}]{Parker_M_L}
{Parker} L.~C.,  {Hudson} M.~J.,  {Carlberg} R.~G.,   {Hoekstra} H.,  2005,
  \mn@doi [\apj] {10.1086/497117}, \href
  {https://ui.adsabs.harvard.edu/abs/2005ApJ...634..806P} {634, 806}

\bibitem[\protect\citeauthoryear{{Peebles} \& {Groth}}{{Peebles} \&
  {Groth}}{1975}]{3PCF_paper}
{Peebles} P.~J.~E.,  {Groth} E.~J.,  1975, \mn@doi [\apj] {10.1086/153390},
  \href {https://ui.adsabs.harvard.edu/abs/1975ApJ...196....1P} {196, 1}

\bibitem[\protect\citeauthoryear{{Schechter}}{{Schechter}}{1976}]{schechter_function_cite}
{Schechter} P.,  1976, \mn@doi [\apj] {10.1086/154079}, \href
  {https://ui.adsabs.harvard.edu/abs/1976ApJ...203..297S} {203, 297}

\bibitem[\protect\citeauthoryear{{Sheldon} et~al.,}{{Sheldon}
  et~al.}{2009}]{Sheldon_M_L}
{Sheldon} E.~S.,  et~al., 2009, \mn@doi [\apj] {10.1088/0004-637X/703/2/2232},
  \href {https://ui.adsabs.harvard.edu/abs/2009ApJ...703.2232S} {703, 2232}

\bibitem[\protect\citeauthoryear{{Spergel} et~al.,}{{Spergel}
  et~al.}{2015}]{wfirst_paper}
{Spergel} D.,  et~al., 2015, arXiv e-prints, \href
  {https://ui.adsabs.harvard.edu/abs/2015arXiv150303757S} {p. arXiv:1503.03757}

\bibitem[\protect\citeauthoryear{{Springel} et~al.,}{{Springel}
  et~al.}{2005}]{Mil_simulation}
{Springel} V.,  et~al., 2005, \mn@doi [\nat] {10.1038/nature03597}, \href
  {https://ui.adsabs.harvard.edu/abs/2005Natur.435..629S} {435, 629}

\bibitem[\protect\citeauthoryear{{Tanimura} et~al.,}{{Tanimura}
  et~al.}{2019}]{warm_hot_gas_filament}
{Tanimura} H.,  et~al., 2019, \mn@doi [\mnras] {10.1093/mnras/sty3118}, \href
  {https://ui.adsabs.harvard.edu/abs/2019MNRAS.483..223T} {483, 223}

\bibitem[\protect\citeauthoryear{{Tegmark} \& {Peebles}}{{Tegmark} \&
  {Peebles}}{1998}]{evolved_bias_model}
{Tegmark} M.,  {Peebles} P.~J.~E.,  1998, \mn@doi [\apjl] {10.1086/311426},
  \href {https://ui.adsabs.harvard.edu/abs/1998ApJ...500L..79T} {500, L79}

\bibitem[\protect\citeauthoryear{{Tegmark} et~al.,}{{Tegmark}
  et~al.}{2004}]{SDSS_fil}
{Tegmark} M.,  et~al., 2004, \mn@doi [\apj] {10.1086/382125}, \href
  {https://ui.adsabs.harvard.edu/abs/2004ApJ...606..702T} {606, 702}

\bibitem[\protect\citeauthoryear{{The Dark Energy Survey Collaboration}}{{The
  Dark Energy Survey Collaboration}}{2005}]{DES}
{The Dark Energy Survey Collaboration} 2005, arXiv e-prints, \href
  {https://ui.adsabs.harvard.edu/abs/2005astro.ph.10346T} {pp
  astro--ph/0510346}

\bibitem[\protect\citeauthoryear{{Vogelsberger} et~al.,}{{Vogelsberger}
  et~al.}{2014}]{Illustris_sim}
{Vogelsberger} M.,  et~al., 2014, \mn@doi [\mnras] {10.1093/mnras/stu1536},
  \href {https://ui.adsabs.harvard.edu/abs/2014MNRAS.444.1518V} {444, 1518}

\bibitem[\protect\citeauthoryear{{Wang} et~al.,}{{Wang}
  et~al.}{2019}]{three_point_correlation_function_prediction}
{Wang} Y.,  et~al., 2019, \mn@doi [\pasa] {10.1017/pasa.2019.5}, \href
  {https://ui.adsabs.harvard.edu/abs/2019PASA...36...15W} {36, e015}

\bibitem[\protect\citeauthoryear{{Wechsler} \& {Tinker}}{{Wechsler} \&
  {Tinker}}{2018}]{relation_galaxy_dmhalo_review}
{Wechsler} R.~H.,  {Tinker} J.~L.,  2018, \mn@doi [\araa]
  {10.1146/annurev-astro-081817-051756}, \href
  {https://ui.adsabs.harvard.edu/abs/2018ARA&A..56..435W} {56, 435}

\bibitem[\protect\citeauthoryear{{Wright}, {Driver}  \& {Robotham}}{{Wright}
  et~al.}{2018}]{wright_M_stellar_M_comparison}
{Wright} A.~H.,  {Driver} S.~P.,   {Robotham} A.~S.~G.,  2018, \mn@doi [\mnras]
  {10.1093/mnras/sty2136}, \href
  {https://ui.adsabs.harvard.edu/abs/2018MNRAS.480.3491W} {480, 3491}

\bibitem[\protect\citeauthoryear{{Xia} et~al.,}{{Xia}
  et~al.}{2019}]{new_filament_paper}
{Xia} Q.,  et~al., 2019, arXiv e-prints, \href
  {https://ui.adsabs.harvard.edu/abs/2019arXiv190905852X} {p. arXiv:1909.05852}

\bibitem[\protect\citeauthoryear{{Yang}, {Mo}, {van den Bosch}, {Pasquali},
  {Li}  \& {Barden}}{{Yang} et~al.}{2007}]{mass_light_ratio_halo}
{Yang} X.,  {Mo} H.~J.,  {van den Bosch} F.~C.,  {Pasquali} A.,  {Li} C.,
  {Barden} M.,  2007, \mn@doi [\apj] {10.1086/522027}, \href
  {https://ui.adsabs.harvard.edu/abs/2007ApJ...671..153Y} {671, 153}

\bibitem[\protect\citeauthoryear{{de Graaff}, {Cai}, {Heymans}  \&
  {Peacock}}{{de Graaff} et~al.}{2019}]{de_graaf_fil}
{de Graaff} A.,  {Cai} Y.-C.,  {Heymans} C.,   {Peacock} J.~A.,  2019, \mn@doi
  [\aap] {10.1051/0004-6361/201935159}, \href
  {https://ui.adsabs.harvard.edu/abs/2019A&A...624A..48D} {624, A48}

\makeatother
\end{thebibliography}
[dataset]* Yang, T., 2020, Data for mass and light map of the filament, Zenodo, \url{https://doi.org/10.5281/zenodo.3951947}

\end{document}